\documentclass[prb,onecolumn,nobibnotes,groupedaddress, natbib]{revtex4}

\usepackage{enumerate}
\usepackage{amssymb}
\usepackage{amsmath}
\usepackage{mathtools}
\usepackage{amssymb}
\usepackage{braket}
\usepackage[hidelinks]{hyperref}
\usepackage{pdfpages}
\usepackage{epstopdf}

\begin{document}


\title{An ``ultimate" coupled cluster method based entirely on $T_2$}

\author{Zachary W. Windom$^{1, 2}$\footnote{\href{mailto:zww4855@gmail.com}{zww4855@gmail.com}},  Ajith Perera$^1$, and Rodney J. Bartlett$^1$}
\affiliation{$^1$Quantum Theory Project, University of Florida, Gainesville, FL, 32611, USA \\
$^2$Computational Sciences and Engineering Division,\ Oak\ Ridge\ National\ Laboratory,\ Oak\ Ridge,\ TN,\ 37831,\ USA}

\begin{abstract}
Electronic structure methods built around double-electron excitations have a rich history in quantum chemistry. However, it seems to be the case that such methods are only suitable in particular situations and are not naturally equipped to simultaneously handle the variety of electron correlations that might be present in breaking a chemical bond, for example. To this end,
 the current work seeks a computationally efficient, low-rank, ``ultimate" coupled cluster method based exclusively on $T_2$ and its products
 which can effectively emulate more ``complete" methods that explicitly consider higher-rank, $T_{2m}$ operators. We introduce a hierarchy of methods designed to systematically account for higher, even order cluster operators - like $T_4, T_6, \cdots, T_{2m}$ - by invoking tenets of the factorization theorem of perturbation theory and expectation-value coupled cluster theory. These higher-order clusters are considered within the coupled cluster doubles (CCD) framework at a manageable impact on calculation cost. It is shown that each member within this methodological hierarchy  is defined such that both the wavefunction and energy are correct through some order in many-body perturbation theory (MBPT), and can be extended up to arbitrarily high orders in $T_2$. The efficacy of such approximations are determined by studying the potential energy surface of several prototypical systems that are chosen to represent both non-dynamic, static, and dynamic correlation regimes. We find that the proposed hierarchy of augmented $T_2$ methods essentially reduce to standard CCD for problems where dynamic electron correlations dominate, but offer improvements in situations where non-dynamic and static correlations become relevant. A notable highlight of this work is that the cheapest methods in this hierarchy - which are correct through fifth-order in MBPT - consistently emulate the behavior of the $\mathcal{O}(N^{10})$ CCDQ method, yet only require a $\mathcal{O}(N^{6})$ algorithm by virtue of factorized intermediates. 

\end{abstract}
\maketitle

\section{Introduction}

Coupled cluster (CC) theory maintains a ``gold-standard" status in computational and quantum chemistry because it routinely provides chemically accurate descriptions of electronic properties\cite{stanton1993equation,bartlett2012coupled} and energetics.\cite{bartlett1981many,bartlett1989coupled} One powerful utility inherent to CC theory is found in its rapid convergence toward the full configuration interaction (FCI) solution, which is the exact answer to the Schrodinger equation within a basis set.  In fact, systematic convergence to the FCI limit can be modulated simply by including higher-order cluster operators.\cite{shavitt2009many} Unfortunately, methods that include full quadruple (CCSDTQ\cite{kucharski1992coupled}), pentuple (CCSDTQP\cite{musial2002formulation}), or hextuple (CCSDTQPH) excitations respectively scale like $\mathcal{O}(N^{10})$, $\mathcal{O}(N^{12})$, and $\mathcal{O}(N^{14})$, thereby limiting these methods' routine use. However, truncated CC approximations - which include only single and double excitation, like CCSD and CCSD(T) - converge so rapidly to the FCI limit that inclusion of higher-order clusters into the ansatz is rarely necessary for chemically useful results, particularly for systems at or near equilibrium geometries.\cite{bartlett2007coupled} Furthermore, every CC approximation guarantees results that are size-extensive.\cite{bartlett1994applications} It is therefore no surprise that CC theory is used across size domains,\cite{windom2022benchmarking,windom2022examining,pavlicek2024comparison,windom2024assessment,mcclain2017gaussian} and is the method of choice to account for dynamic correlations when both high-accuracy and computational tractability are required.

However, in situations where non-dynamic correlations become significant - as in the breaking of multiple bonds - truncated CC approximations can experience difficulty converging toward the FCI limit. Indeed, there are several examples illustrating poor or even divergent convergence behavior, with a famous example being the dissociation of the triply bonded nitrogen dimer.\cite{krogh2001general,kucharski1999geometry} In this case, a conventional viewpoint might assert that the CC ansatz should either include higher-order cluster operators or adopt a multireference starting point\cite{li1998dissociation,laidig1987description} to improve upon commonly used, truncated CC approximations. This is quite the conundrum, as the standard route of including higher-order clusters as previously alluded to can become quite expensive, and the developed of multireference CC approaches are not yet mature enough to be packaged as a computational ``blackbox." 

One way to circumnavigate the addition of expensive, higher-order terms to the CC ansatz involves throwing out particular diagrams known to cause problems upon dissociation. This is the philosophy of "Addition by subtraction" in CC theory,\cite{bartlett2006addition,musial2007addition,rishi2019behind} and is a guiding principle behind the distinguishable cluster approximation to coupled cluster with single and double excitations (DCSD), which has shown success in instances of static correlation.\cite{kats2013communication,kats2014communication,rishi2016assessing} This viewpoint is taken to the extreme by the pair coupled cluster with double excitations (pCCD) method,\cite{cullen1996generalized} wherein all off-diagonal elements of the double excitation tensor, $T_2$, are zero  ($\forall$ $i\neq j$, $a \neq b$, $t_{ij}^{ab}=0$). In spite of the significant simplification to $T_2$, pCCD has shown some promise\cite{henderson2014seniority,kossoski2023seniority,kossoski2021excited,boguslawski2014efficient,brzek2019benchmarking,boguslawski2017benchmark,henderson2015pair,nowak2021orbital} all the while exhibiting an attractive $\mathcal{O}(N^3)$ scaling. Rationale behind pCCD's success can be found in its association with doubly occupied CI (DOCI), which can be thought of as a constrained FCI that only exploits the senority zero sector of Hilbert space.\cite{stein2014seniority} The ``perfect pairing" hierarchy of methods was also proposed to similarly constrain a set of cluster operators to satisfy a generalized valence bond type of pairing scheme, leading to methods that are cheap compared to their standard CC alternatives.\cite{parkhill2010truncation,lehtola2018orbital,lehtola2016cost}

Instead of throwing terms out of the CC ansatz, an alternative approach can be conceived that focuses on the systematic improvement to $T_2$, and might include at least partial support for higher-order excitation effects through $T_{2n}$. 
This work is primarily concerned with methods that satisfy these prerequisites because electron correlation effects are initially accounted for in first-order via $T_2$ and the process of breaking $n$ bonds can be understood to involve $2n$ pairs of electrons. To this end, the baseline CC model we are focused on systematically improving is the CCD ansatz. This is because the CCD ansatz is among the cheapest CC methods that is capable of accounting for quadruple excitations via the connected product, $T_2^2$. However, the variational or expectation-value CCD (VCCD or XCCD, respectively) method might be viewed as a natural asymptote for a purely $T_2$ based method 
since the energy is expressed in terms of an expectation value and is variationally optimized. Assuming the $T_2$ amplitudes are obtained by invocation of the Rayleigh-Ritz variational principle, VCCD/XCCD would represent an upper bound to the FCI which would have obvious benefits when studying cumbersome portions of a potential energy surface.\cite{marie2021variational,van2000benchmark,evangelista2011alternative} However, it should be noted that both FCI and VCCD show similar scalings\cite{van2000benchmark} and are therefore intractable beyond small model systems. These scaling issues have been addressed to some extent in recent implementations of VCC which exploit a FCI program, and have show improvements over CCD in instances of non-dynamic correlation.\cite{robinson2012quasi,robinson2012benchmark} Similar work on quadratic CCD (QCCD) - proposed as a tractible intermediate between CCD and extended CC (ECC\cite{arponen1983variational,arponen1987extended}) - has also shown promise.\cite{van2000quadratic,gwaltney2002perturbative}

The current work seeks alternative avenues for designing tractable approximations to an ultimate $T_2$ method. To this end we explore two independent strategies: the first builds upon early efforts based on XCC,\cite{bartlett1988expectation,bartlett1989some,bartlett2024perspective} while the second invokes arguments based on the CC functional to derive corrections to the energy based on orders of perturbation theory.\cite{kucharski1998sixth,taube2008improving} We will show both viewpoints are intimately connected to work that utilizes the factorization theorem,\cite{frantz1960many,bartlett1975some} which laid the groundwork for later, affordable methods designed to estimate the effects of connected quadruple excitations.\cite{kucharski1986fifth,kucharski1989coupled,kucharski1993coupled,kucharski1998noniterative,kucharski1999connected,kucharski1998efficient,musial2010improving,thorpe2024factorized} This work explores the possibility of accounting for higher-order cluster effects associated with $T_{2n}$ using only $T_2$ and its adjoint. We show the merits of this approach, and further emphasize the improvements in the computational scaling that can be made over a corresponding ``full" CC method which might otherwise explicitly account for $T_{2n}$ in the iterative solution of the residual equations.

The rest of the paper is organized as follows. The Theory section discusses the background relevant to the current paper, and points to the Appendix wherein useful derivations have been provided that discuss alternative approaches which might be conceived. The Theory section is followed by the Computational Details section, after which comparison of the proposed methods are benchmarked in the Results subsection. Concluding remarks then follow.

\section{Theory}
Standard coupled cluster theory starts with the exponential ansatz acting on some single-reference - in this case, the Hartree-Fock - determinant, $\ket{0}$.
\begin{equation}\label{eq:ccwvn}
    \Psi_{CC}=e^{T}\ket{0}
\end{equation} The cluster operator, $T$, can be decomposed into contributions associated with single, double, triple, etc excitations
\begin{equation}
    T=T_1+T_2+T_3+\cdots
\end{equation}, where each $T_n$ operator can be expressed in second quantization
\begin{equation}
    T_n=\frac{1}{(n!)^2}\sum_{occ,virt} t_{ijk\cdots}^{abc\cdots}\{a^{\dag}i b^{\dag}jc^{\dag}k \cdots \}
\end{equation} and the sum is over occupied ($i,j,k,l,\cdots$) and virtual ($a,b,c,d,\cdots$) spin-orbitals. We adopt the standard notation that assumes unrestricted indices $p,q,r,s,\cdots$. Using these labeling conventions, we define the normal-ordered Hamiltonian as
\begin{equation}
\begin{split}
    H_N = & H - \braket{0|H|0} \\
    =&\sum_{p.q}f_{pq}\{p^{\dag}q\} +\frac{1}{4}\sum_{p,q,r,s}\braket{pq||rs}\{p^{\dag}q^{\dag}sr\} \\
        =   & f_N + W_N
    \end{split}
\end{equation} Standard CC theory then gives the correlation contribution as 
\begin{equation}\label{eq:corrE}
    E_{CC} = \braket{0|\bigg( H_Ne^T\bigg)_c|0}
\end{equation} using amplitudes recovered by solving residual equations of the form
\begin{equation}\label{eq:residEqns}
    Q\bigg( H_Ne^T\bigg)_c\ket{0} = 0
\end{equation} where $Q=I-\ket{0}\bra{0}=Q_1+Q_2+\cdots$  is the complement to the single-reference and serves the purpose of projecting these equations onto a particular excitation manifold. The subscript $c$ denotes the restriction to connected diagrams.

The current work begins by restricting $T$ to $T_2$ and its (disconnected) products.  As our purpose in this work is to explore ``ultimate" $T_2$ methods that systematically improve upon the standard CCD model, refinements to the energy and/or amplitude expressions are sought outside the traditional confines of Equations \ref{eq:corrE} and \ref{eq:residEqns}. Our objective is to extract principal effects of higher-order connected clusters, like $T_4$, $T_6$, etc by evaluating products of $T_2$. We outline two complementary ways of achieving such an ``ultimate" $T_2$ method. The first - based on XCCD - is discussed in the following section. A related approach - based on the $\Lambda$ functional - is discussed in the Appendix and is shown to yield identical energy corrections at low (5th, 6th, and 8th) orders if all energy diagrams are capped using two-electron integrals. The relationship between the factorization theorem of MBPT and the $\Lambda$-based approach is reviewed, and we further show that within the confines of the current work, both approaches yield equivalent energy diagrams. We show that this is a direct consequence of both approaches eliminating the highest rank denominator in favor of a product string of $D_2$ denominators.

 \clearpage

\subsection{XCCD(k) route toward an ultimate T2 method}

The untruncated expectation-value coupled cluster doubles theory is formulated by  expressing the correlation energy as
\begin{equation} \label{eq:expCC}
\begin{split}
    \Delta E=&\frac{\braket{0|e^{T_2^{\dag}}H_Ne^{T_2}|0}}{\braket{0|e^{T_2^{\dag}}e^{T_2}|0}} \\
    =&\braket{0|\big(e^{T_2^{\dag}}H_Ne^{T_2}\big)_C|0}\\
    \end{split}
\end{equation} which is strictly limited to connected diagrams, analogous to standard CC theory.

From an XCC-viewpoint, an energy functional can be constructed to coincide with an even or odd order in MBPT.\cite{bartlett1989some} For example, we find that at even orders, $k=2n$, in MBPT the corresponding correction is
\begin{equation}\label{eq:xcceven}
    \Delta E^{[2n]} =\braket{0|\frac{\big(T_2^{\dagger}\big)^{n}}{n!}f_N\frac{\big(T_2\big)^n}{n!}|0} + \bigg(\braket{0|\frac{\big(T_2^{\dagger}\big)^{n}}{n!}W_N\frac{\big(T_2^{\dagger}\big)^{n-1}}{(n-1)!} |0} +\text{h.c.}\bigg)
\end{equation} where we note that $h.c.$ stands for the hermitian conjugate of the preceding term. On the other hand, at odd orders, $k=2n+1$, we find that

\begin{equation}\label{eq:xccodd}
    \Delta E^{[2n+1]}=\braket{0|\frac{\big(T_2^{\dagger}\big)^{n}}{n!}W_N\frac{\big(T_2\big)^n}{n!}|0}
\end{equation} Independent of the order, only fully linked contributions of the above can contribute  to the correlation energy. Note that the superscript here (e.g. $k=2n$ or $k=2n+1$) denotes an individual contribution that arises at a particular order in MBPT. If we wanted to determine the complete correlation correction thru some order in MBPT, we need to sum all prior, individual contributions such that
\begin{equation}\label{eq:infEexp}
    \Delta E(k) = \sum_{k=2}^{\infty}\Delta E^{[k]}
\end{equation}

  It should be emphasized that when working out derivations for higher-order equations, simplifications arise from isolating and eliminating lower-order XCCD(k) residual equations that may be embedded in the current working expressions. We follow this philosophy in the following derivations of the XCCD(2-6), and then provide general equations that describe the XCCD residual equations and simplified energy functional at arbitrary order, $k$. By taking this expansion to infinite-order, the resulting  XCCD($k\rightarrow \infty$) method should asymptotically approach a variational CCD limit.

\subsubsection{XCCD(2)}
At second order in MBPT, we find that 
\begin{equation}\label{eq:XCCD2energy}
    \Delta E^{[2]} = \braket{0|T_2^{\dag}f_NT_2|0}_c + \bigg( \braket{0|T_2^{\dag}W_n|0}+h.c.\bigg)
\end{equation} By varying Equation \ref{eq:XCCD2energy} with respect to $T_2^{\dag}$, the corresponding set of residual equations are recovered as
\begin{equation}\label{eq:XCCD2resid}
    Q_2\bigg( f_NT_2 + W_N\bigg)_c\ket{0}=0
\end{equation} which is equivalent to the MBPT(2) residual equations. By invoking the residual condition of Equation \ref{eq:XCCD2resid} on Equation \ref{eq:XCCD2energy}, we see that the correlation energy reduces to
\begin{equation}\label{eq:trueXCCD2energy}
   \Delta E^{[2]} = \braket{0|(W_NT_2)_c|0}  
\end{equation}

\subsubsection{XCCD(3)}
The individual third order contribution appears as
\begin{equation}\label{eq:lccd}
    \Delta E^{[3]} = \braket{0|T_2^{\dag}W_NT_2|0}_c
\end{equation}. Using Equations  \ref{eq:infEexp}, \ref{eq:XCCD2energy}, and \ref{eq:lccd} variation of $\Delta E(3) = \Delta E^{[3]} + \Delta E^{[2]}$ with respect to $T_2^{\dag}$ yields the corresponding residual equation
\begin{equation}\label{eq:XCCD3resid}
    Q_2\bigg(f_NT_2 + W_N + W_NT_2 \bigg)_c\ket{0} = 0
\end{equation} which are seen to be equivalent to those of LCCD. Incorporating Equation \ref{eq:XCCD3resid} back into the $E(3)$ energy expression simplifies to the same expression as in Equation \ref{eq:trueXCCD2energy}. Regardless, this energy is evaluated using comparably improved LCCD amplitudes and is correct through third-order in MBPT.

\subsubsection{XCCD(4)}
At fourth-order, we have 
\begin{equation}\label{eq:startXCCD4energy}
    \Delta E^{[4]} = \braket{0|\frac{T_2^2}{2}^{\dag}\bigg(f_N\frac{T_2^2}{2}\bigg)_D|0}_c +\braket{0|\frac{T_2^2}{2}^{\dag}\bigg(W_NT_2\bigg)_D|0}_c + \braket{0|T_2^{\dag}W_N\frac{T_2^2}{2}|0}_c
\end{equation} where the $D$ denotes internally disconnected diagrams that arise for the first time. At this order and beyond, internally disconnected terms will appear in the energy functional which will need to be canceled before the residual equations are constructed if the overall, finite-order theory is to remain fully linked, and by extension, size-extensive. The way this is done henceforth is by invoking lower-order residual equations to ``zero-out" the disconnected terms' contribution. Alternatively, we could assert a proposition wherein all internally disconnected terms should automatically be eliminated \textit{a priori}. We adopt the more rigorous, former option in subsequent derivations.

 Recognizing that the first two terms in Equation \ref{eq:startXCCD4energy} are composed of internally disconnected diagrams despite being connected \textit{en toto}, we can manipulate these terms such that
\begin{equation}\label{eq:cancelXCCD4}
    \braket{0|\frac{T_2^2}{2}^{\dag}T_2Q_2\bigg(f_NT_2+W_N\bigg)_c|0}=0+\delta(5)
\end{equation} where the resulting quantity vanishes up to fifth-order terms, $\delta(5)$. This can be rationalized by recognizing $\big( f_NT_2+W_N\big)_C$ is 0 by virtue of the XCCD(2) equations (Equation \ref{eq:XCCD2resid}), and each additional $T_2$/$T_2^{\dagger}$ beyond that increases the order of the error by one. In the current situation where we have limited ourselves to truncations of the XCCD energy functional at fourth-order in MBPT, these two terms may be eliminated. Now that we have eliminated the internally disconnected terms found in Equation \ref{eq:startXCCD4energy}, we can now express the cumulative fourth-order energy functional as
\begin{equation}\label{eq:fullXCCD4energy}
    \Delta E(4)=\braket{0|T_2^{\dag}W_N\frac{T_2^2}{2}|0}_c + \braket{0|T_2^{\dag}W_NT_2|0}_c +\braket{0|T_2^{\dag}f_NT_2|0}_c + \braket{0|T_2^{\dag}W_N|0}_c + \braket{0|W_NT_2|0}_c
\end{equation} using contributions from Equations \ref{eq:XCCD2energy}, \ref{eq:lccd}, and the internally connected part of \ref{eq:startXCCD4energy}.

Upon variation of Equation \ref{eq:fullXCCD4energy} w.r.t $T_2^{\dag}$, we recover the following residual equations
\begin{equation}\label{eq:XCCD4resid}
    Q_2\bigg( W_N\frac{T_2^2}{2} + W_NT_2 + f_NT_2 + W_N\bigg)_c\ket{0}=0
\end{equation} Inserting this residual condition back into Equation \ref{eq:fullXCCD4energy} leads to the XCCD(4) energy that, again, appears as Equation \ref{eq:trueXCCD2energy}. Note that the XCCD(4) equations are identical to those found in standard CCD.

\subsubsection{XCCD(5)}
At fifth-order we notice that the linked energy expression can be decomposed into internally disconnected (D) and connected (c) contributions seen by
\begin{equation}\label{eq:startXCCD5energy}
\begin{split}
        \Delta E^{[5]} &= \braket{0|\big(\frac{T_2^2}{2}\big)^{\dag}W_N\frac{T_2^2}{2}  |0}_c \\
        &= \braket{0|\bigg(\frac{T_2^2}{2}\bigg)^{\dag}\bigg(\frac{W_NT_2^2}{2}\bigg)_D |0}_c +\braket{0|\bigg(\frac{T_2^2}{2}\bigg)^{\dag}\bigg(\frac{W_NT_2^2}{2}\bigg)_c |0}_c
\end{split}
\end{equation} After combining the disconnected term in Equation \ref{eq:startXCCD5energy} with those found at fourth order (see  Equations \ref{eq:startXCCD4energy} and \ref{eq:cancelXCCD4}), we find that all internally disconnected terms can be cancelled 

\begin{equation}\label{eq:XCCD5discon}
\begin{split}
\braket{0|\bigg(\frac{T_2^2}{2}\bigg)^{\dag} \bigg( W_N\frac{T_2^2}{2} + f_N\frac{T_2^2}{2} +W_NT_2\bigg)_D|0}&=\braket{0|\bigg(\frac{T_2^2}{2}\bigg)^{\dag}T_2Q_2\bigg(W_NT_2+W_N+f_NT_2\bigg)_c|0} \\
&=0+\delta(6)
\end{split}
\end{equation} through sixth-order in MBPT. Here, we have exploited the XCCD(3) residual equations of Equation \ref{eq:XCCD3resid} which are embedded within Equation \ref{eq:XCCD5discon} to facilitate this cancellation. Hence, these terms can be eliminated through sixth-order. Consequently, the cumulative correlation correction through fifth-order is shown to be
\begin{equation}\label{eq:fullXCCD5energy}
\begin{split}
    \Delta E(5) = &\braket{0|\bigg(\frac{T_2^2}{2}\bigg)^{\dag}\bigg(\frac{W_NT_2^2}{2}\bigg)_c |0}_c\\
     &         + \braket{0|T_2^{\dag}\bigg(\frac{W_NT_2^2}{2} + W_NT_2 +f_NT_2 + W_N \bigg)_c|0}_c \\
     &         + \braket{0|\bigg(W_NT_2\bigg)_c|0}_c
    \end{split}
\end{equation} where we organize the expression to enhance the clarity of each terms' dependence on powers of $T_2^{\dag}$. We now vary Equation \ref{eq:fullXCCD5energy} w.r.t. $T_2^{\dag}$ to determine the residual equations for XCCD(5)
\begin{equation}
    Q_2\bigg( T_2^{\dag}\big(\frac{W_NT_2^2}{2}\big)_c + \big(\frac{W_NT_2^2}{2}\big)_c + \big(W_NT_2\big)_c + \big(f_NT_2\big)_c +W_N\bigg)=0
\end{equation} By enforcing the residual condition back on Equation \ref{eq:fullXCCD5energy}, we note that while the second line completely cancels, the first line of Equation  \ref{eq:fullXCCD5energy} is incompletely canceled while the third line is left untouched. Therefore, the simplified energy functional of XCCD(5) in the presence of converged amplitudes can be expressed as 
\begin{equation}\label{eq:XCCD5finalenergy}
    \Delta E(5) = \braket{0|(W_NT_2)_c|0}_c - \braket{0|\bigg(\frac{T_2^2}{2}\bigg)^{\dag}\bigg(\frac{W_NT_2^2}{2}\bigg)_c |0}_c
\end{equation}

\subsubsection{XCCD(6)}
The sixth order energy expression is shown to be
\begin{equation}\label{eq:startXCCD6energy}
    \Delta E^{[6]} = \braket{0|\bigg(\frac{T_2^3}{3!}\bigg)^{\dag}\bigg( \frac{f_NT_2^3}{3!}\bigg)_D|0}_c + \braket{0|\bigg(\frac{T_2^3}{3!}\bigg)^{\dag}\bigg( \frac{W_NT_2^2}{2!}\bigg)_D|0}_c + \braket{0|\bigg(\frac{T_2^2}{2!}\bigg)^{\dag}\bigg( \frac{W_NT_2^3}{3!}\bigg)|0}_c
\end{equation} where - as in XCCD(4) - we are immediately interested in eliminating the two internally disconnected diagrams of Equation \ref{eq:startXCCD6energy}. In analogy to the procedure followed in XCCD(4) in this situation, we can extract a factor of $\bigg(\frac{T_2^3}{3!}\bigg)^{\dag}\frac{T_2^2}{2}$ from both disconnected terms in the above to find the embedded XCCD(2) residual equation. Upon doing this, we find that these two terms in  Equation \ref{eq:startXCCD6energy} cancel through $\delta(7)$. 

Paying close attention to the third term in Equation \ref{eq:startXCCD6energy}, we note that this term can also be decomposed into internally connected and disconnected contributions such that the overall diagram is connected. Consequently, we have to consider how to handle the internally disconnected portion of
\begin{equation}\label{eq:decomposeXCCD6}
    \braket{0|\bigg(\frac{T_2^2}{2!}\bigg)^{\dag}\biggl\{\bigg(\frac{W_NT_2^3}{3!}\bigg)_D + \bigg(\frac{W_NT_2^3}{3!}\bigg)_c\biggr\}|0}_c
\end{equation} Referring back to Equation \ref{eq:XCCD5discon}, we note that a natural continuation of the sequence can proceed through sixth order after including the internally disconnected part of Equation \ref{eq:decomposeXCCD6} such that
\begin{equation}\label{eq:XCCD6seq}
    \braket{0|\bigg(\frac{T_2^2}{2}\bigg)^{\dag}\biggl\{ \frac{W_NT_2^3}{3!} + W_N\frac{T_2^2}{2} + f_N\frac{T_2^2}{2} +W_NT_2 \biggr\}|0}
\end{equation} Noting that after we extract a factor of $T_2$ from the above expression, we find the XCCD(4) residual equations (see Equation \ref{eq:XCCD4resid}) embedded within Equation \ref{eq:XCCD6seq}. Consequently, this  facilitates elimination of these terms with error $\delta(7)$. 

Now, we are left with a purely internally connected piece in $\Delta E^{[6]}$, which is added on top of $\Delta E(5)$ in Equation \ref{eq:fullXCCD5energy} to form the sixth order energy functional, $\Delta E(6)$. We can then form the corresponding set of residual equations by taking a partial derivative of this quantity to form


\begin{equation}
\begin{split}
      Q_2\bigg(T_2^{\dag}\biggl\{ \frac{W_NT_2^3}{3!} + \frac{W_NT_2^2}{2}\biggr\}_c + \biggl\{ \big(\frac{W_NT_2^2}{2}\big)_c + \big(W_NT_2\big)_c + \big(f_NT_2\big)_c +W_N\biggr\}_c \bigg) =0  
\end{split}
\end{equation} Enforcing the stationarity of these equations back on the original energy functional, we find that the final, simplified sixth-order energy is 
\begin{equation}
    \Delta E(6)=\braket{0|(W_NT_2)_c|0}_c - \braket{0|\bigg(\frac{T_2^2}{2}\bigg)^{\dag}\bigg(\frac{W_NT_2^2}{2}\bigg)_c |0}_c - \braket{0|\bigg( \frac{(T_2^2)^{\dagger}}{2}\bigg)\bigg(\frac{W_NT_2^3}{3!}\bigg)_c|0}_c
\end{equation}

\subsubsection{XCCD(k), for arbitary $k$}

The generic procedure to build arbitrary order XCCD working equations can be informed by the prior derivations. They can be briefly summarized in the following three steps:

\begin{enumerate}
    \item Build the complete energy functional using Equations \ref{eq:xcceven}, \ref{eq:xccodd}, and \ref{eq:infEexp}. Simplify this result by cancelling all internally disconnected components of the energy functional; this is supported by the induction proof provided in the Appendix  which verifies that the only new additions to the energy functional at even orders, $k=2n$ is
    \begin{equation}\label{eq:generalevenXCC}
        \Delta E^{[2n]}= \braket{0|\bigg( \frac{(T_2^{\dagger})^{n-1}W_NT_2^n}{(n-1)!n!}\bigg)_c|0}
    \end{equation} whereas at odd orders, $k=2n+1$,
    \begin{equation}\label{eq:generaloddXCC}
        \Delta E^{[2n+1]}=\braket{0|\bigg( \frac{(T_2^{\dagger})^nW_NT_2^n}{n!n!}\bigg)_c|0}
    \end{equation}  

    \item After the starting energy functional has been simplified by elimination of all internally disconnected components, take the partial derivative to generate the corresponding set of residual equations. Using the prior two equations, we find that at arbitrary even orders, $k=2n$, the term 
        \begin{equation}\label{eq:evenxccgeneralreside}
        Q_2\bigg( \frac{(T_2^{\dagger})^{n-2}}{(n-2)!}W_N\frac{T_2^n}{n!}\bigg)_C\ket{0}
    \end{equation}
 is added while at arbitrary odd orders, $k=2n+1$, the term

     \begin{equation}\label{eq:oddxccgeneralreside}
        Q_2\bigg( \frac{(T_2^{\dagger})^{n-1}}{(n-1)!}W_N\frac{T_2^n}{n!}\bigg)_C\ket{0}
    \end{equation}
 is added.

    \item Invoke the residual condition of Equations \ref{eq:evenxccgeneralreside} and \ref{eq:oddxccgeneralreside} on the simplified energy functional which has been constructed with respect to the addition of terms in Equations \ref{eq:generalevenXCC} and \ref{eq:generaloddXCC}. Upon doing so, we find that the final form of the energy functional appears as
    \begin{equation}\label{eq:finaleveneqn}
        \Delta E^{[2n]}=\frac{1}{n!}\bigg(\frac{1}{(n-1)!} - \frac{1}{(n-2)!}\bigg)\braket{0|(T_2^{\dagger})^{n-1}W_NT_2^{n}|0}
    \end{equation}
    and 
    \begin{equation}\label{eq:finaloddeqn}
        \Delta E^{[2n+1]}=\frac{1}{n!}\bigg(\frac{1}{n!}-\frac{1}{(n-1)!} \bigg)\braket{0|(T_2^{\dagger})^nW_nT_2^n|0}
    \end{equation} These two equations, in conjunction with Equation \ref{eq:infEexp}, can be used to recursively generate the final XCCD(k) energy functional through arbitrary order $k>4$. 
\end{enumerate} In summary, Equations \ref{eq:evenxccgeneralreside} and \ref{eq:oddxccgeneralreside} can be used to recursively define the residual equations for a method at arbitrary order, whereas Equations \ref{eq:finaleveneqn} and \ref{eq:finaloddeqn} would be used to recursively define the final energy expression that corresponds to the method at arbitrary order. These four equations are all that is needed to build XCCD(k).

\subsection{Computational complexity of $T_2$-only strategies}

If we were interested in explicitly coupling the $T_4$ or the $T_4$ and $T_6$ operators to the $T_2$ equations, the standard CC approach would lead to the CCDQ and CCDQH methods which scale as 
$\mathcal{O}(N^{10})$ and $\mathcal{O}(N^{14})$ respectively. Unfortunately, the steep, high-polynomial scaling of these methods arises in part from contractions between the $T_4$/$T_6$ cluster operators and their corresponding $D_4$/$D_6$ Fock energy denominator. If a procedure was in place to  eliminate the need to contract against these denominators, then ``factorized" intermediates could be constructed that reduce the methods' algorithmic complexity.

\begin{figure}
\includegraphics[scale=0.25]{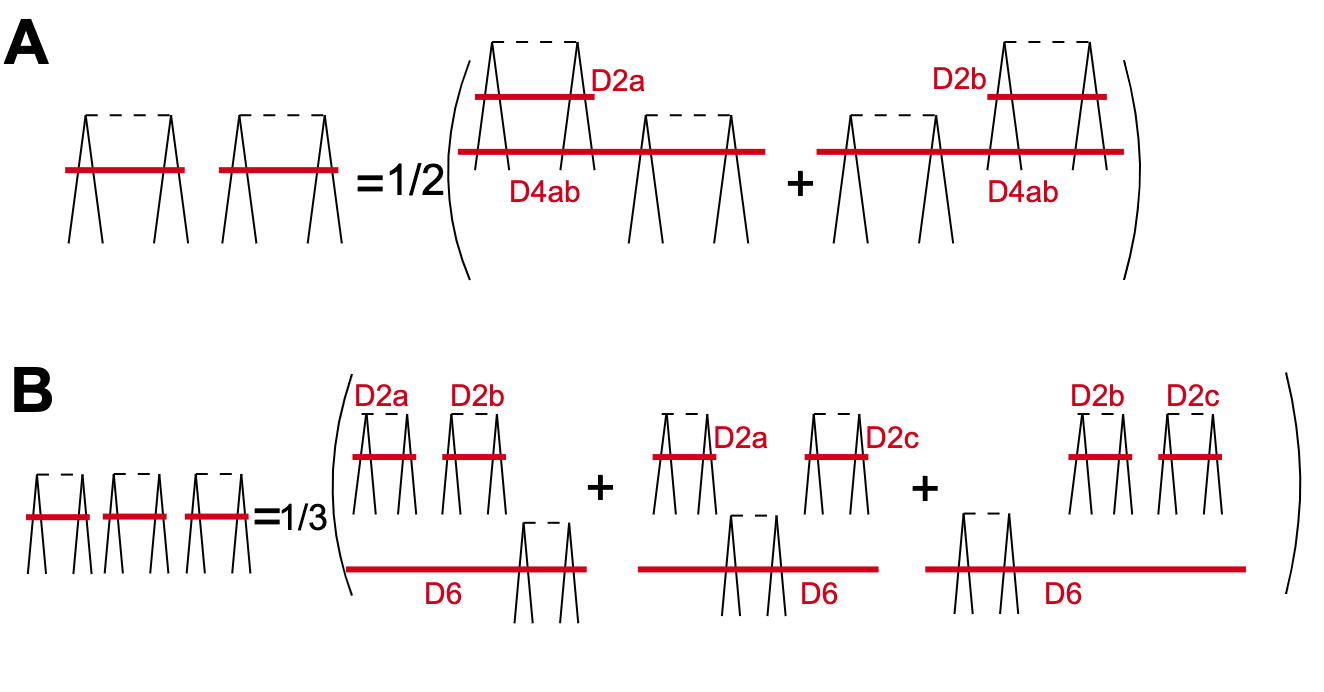}
\caption{Depiction of the factorization theorem for A) quadruples and B) hextuples. We associate each solid red line crossing 4 particle/hole lines (on LHS) with a $D_2$ denominator; in the case of it crossing 8 particle/hole lines, we associated a $D_4$ denominator, and so on. Note that the RHS also labels  the particular time-orderings with an  $a$, $b$, or $c$ to denote the distinct time-orderings that are possible. It is important to recognize that the factorization employed in A) was used to to eliminate the $D_4$ denominators in B), leading to a common time-ordering for two of the three vertices in each term. Factor of $\frac{1}{2}$ and $\frac{1}{3!}$ implicit on the LHS of A) and B), respectively. }
\label{fig:factorizationTheorem}
\end{figure}

There are two  techniques that could conceivably advance such a philosophy. The first family of approaches involves tracing higher-rank, CCDQ/CCDQH residual equations to determine their effect on the energy, and then invoke the factorization theorem of MBPT\cite{frantz1960many} to explicitly eliminate higher-rank denominators. In the context of our current objective, Figure \ref{fig:factorizationTheorem} demonstrates how the factorization theorem can be utilized to eliminate $D_4$ and $D_6$ denominators in favor of a product string of $D_2$ denominators. Similar strategies have been used previously in the Bartlett group with much success.\cite{kucharski1986fifth,kucharski1998efficient,musial2000t5,musial2001coupled} The second is based on XCCD, which implicitly supports higher-rank cluster operator effects, but without any contraction against higher-rank denominators by construction. This can be seen, for example, from the relationship between $T_4$ and $T_6$ to the highest-order terms in XCCD(5/6) and XCCD(7-9), respectively, and is discussed to some extent in the Appendix and in more detail elsewhere.\cite{strongCorrT2}

Although there are slight differences between these approaches, the fact that higher-rank energy denominators have been eliminated supports the construction of factorized intermediates. A more in-depth discussion of these two techniques and their associated scaling is presented in prior work.\cite{kucharski1986fifth,strongCorrT2} Nevertheless, we note that in the context of the current work that the highest order terms in XCCD(5) and XCCD(8) - which implicitly emulate $T_4$ and $T_6$ excitation effects, respectively - can be constructed using algorithms that scale like $\mathcal{O}(N^{6})$ and $\mathcal{O}(N^{10})$, respectively. In other words, XCCD(5) has an algorithmic complexity that is equivalent to CCD yet the former has implicit support for quadruple excitations. Similarly, XCCD(8) has an algorithmic complexity that is equivalent to CCDQ yet offers implicit support for hextuple excitations. This establishes a potential route to circumnavigate the $\mathcal{O}(N^{10})$ and $\mathcal{O}(N^{14})$ scaling of CCDQ and CCDQH, respectively, using methods that nevertheless implicitly support higher-rank excitation effects.

\clearpage

\section{Computational Details}
\begin{figure}
\includegraphics[scale=0.15]{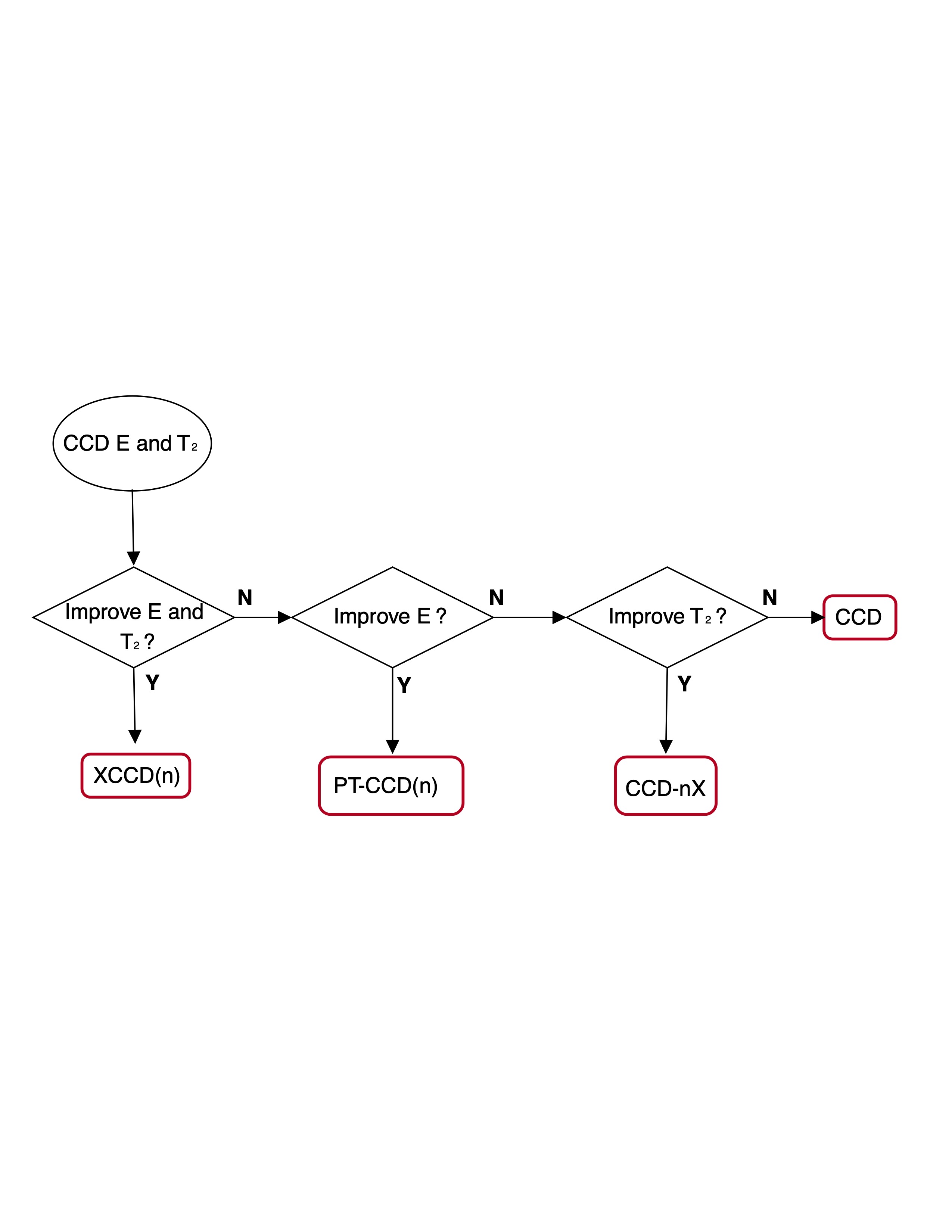}
\caption{Flowchart depicting the hierarchy of methods which are achievable by customizing corrections to standard CCD. Our choices are to focus our efforts on improving either the CCD energy, the $T_2$ ansatz, or both at the same time (eg. XCCD).}
\label{fig:XCCflowchart}
\end{figure}

\begin{figure}
\includegraphics[scale=0.15]{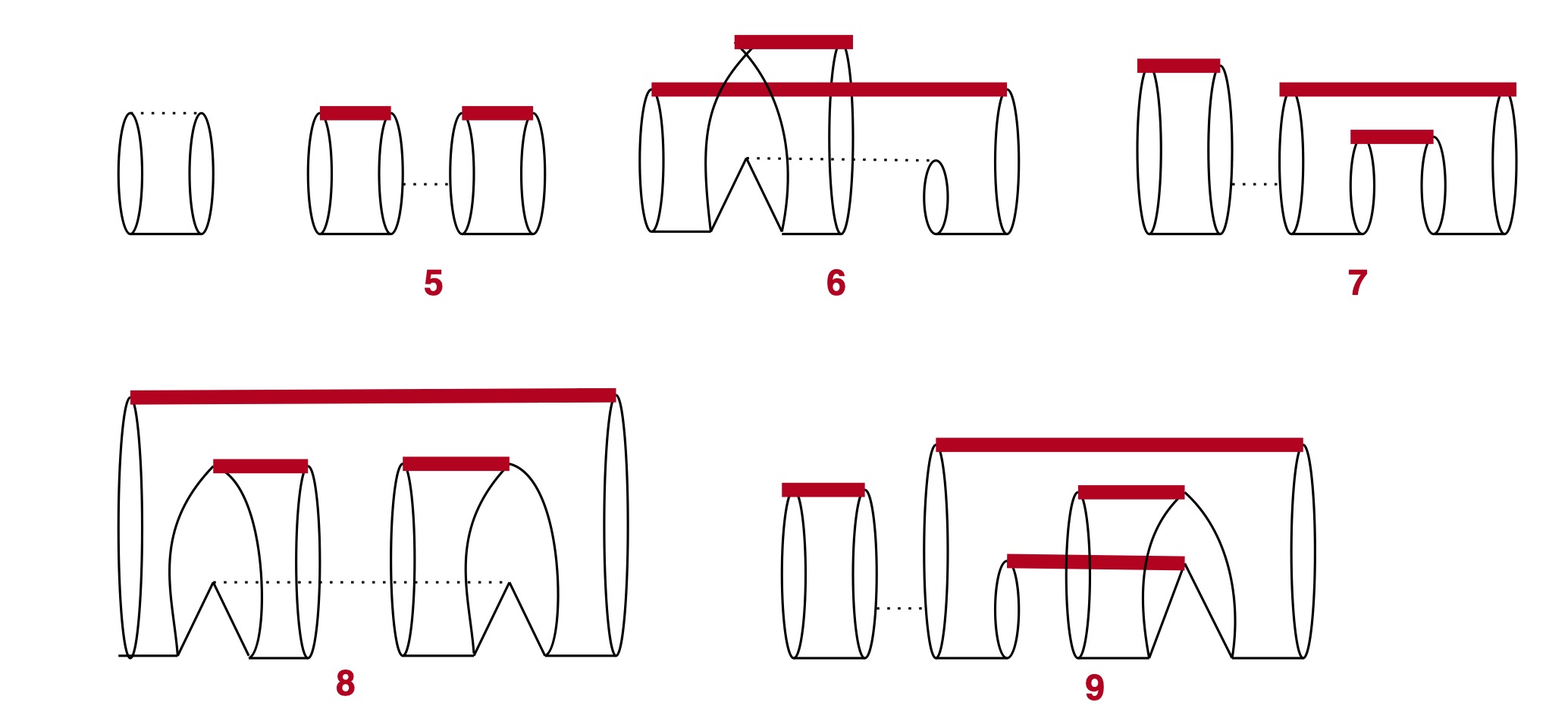}
\caption{Skeleton diagrams of the XCCD($k$) and PT-CCD($k$) energy expressions, for $k=5-9$. For both approaches, the red cap denotes a $T_2^{\dagger}$ vertex. It should be noted that the XCCD(2-4) energy consists of only the first diagram, $\braket{0|W_NT_2|0}$. }
\label{fig:XCCDenergy}
\end{figure}

We explore 3 flavors of approximation that have been informed by our XCCD($k$) developments. The particular methods explored in this work each focuses on either perturbative improvements to the CCD energy, the CCD ansatz, or both (e.g. the pure XCCD($k$) approach). The resulting hierarchy of methods are summarized below:
 \begin{itemize}
 \item \textbf{PT-CCD($k$)} Augments the energy of a converged CCD calculation by adding perturbative energy corrections that have been informed by the cumulative XCCD($k$) energy functional
 \item \textbf{CCD-$k$X} Augments the CCD $T_2$ residual equations by including higher-order diagrams found in the XCCD($k$) residual equations
\item \textbf{XCCD($k$)}: Augments both the CCD energy and wavefunction according to the previously derived equations
\end{itemize}
The algorithmic workflow differentiating these methods - with respect to each other and the standard CCD method - is neatly summarized in the flowchart  of Figure \ref{fig:XCCflowchart}. In particular, we explore methods that are complete through orders $k=5-9$;  as an example, Figure \ref{fig:XCCDenergy} depicts the energy diagrams that show up thru order $k=9$. Regarding the augmentation of the $T_2$ residual equations, the CCD-$k$X and XCCD($k$) methods include the $T_2$ base of the diagrams that have been capped with $(n-1)$ $T_2^{\dagger}$'s.  


The previously discussed methods were implemented in a custom Python package, UT2, which relies on PySCF\cite{sun2018pyscf} for SCF capabilities as well as 1 and 2 body integrals. Construction of higher-order diagrams was supported by automatic diagram generating software; notably $p^{\dagger}q$\cite{rubin2021p} and Wicked.\cite{evangelista2022automatic} The UT2 implementation for the lowest-order, perturbative corrections were verified to machine precision using the CCSDT(Qf)\cite{kucharski1998noniterative}  capabilities of ACESII.\cite{perera2020advanced} Checks against other methods\cite{kucharski1998efficient} were also made.  

This work focuses on the ground state electronic structure of C$_2$H$_2$, N$_2$,  H$_2$O, and the H$_4$ ring. These systems were chosen as representatives for different types of electron correlation: dynamic, non-dynamic, and static. Of particular interest in the current work is the ability of the aforementioned methods to describe non-dynamic correlation. As has been pointed out, minimal basis sets accentuate this type of correlation.\cite{van2000benchmark}  Consequently, the STO-6G basis set is used, although we also include results using the DZ and DZP basis sets when applicable. Unless otherwise specified, all calculations use a RHF reference. 

\section{Results}

\subsection{H$_2$O}

\begin{figure}
\includegraphics[scale=0.15]{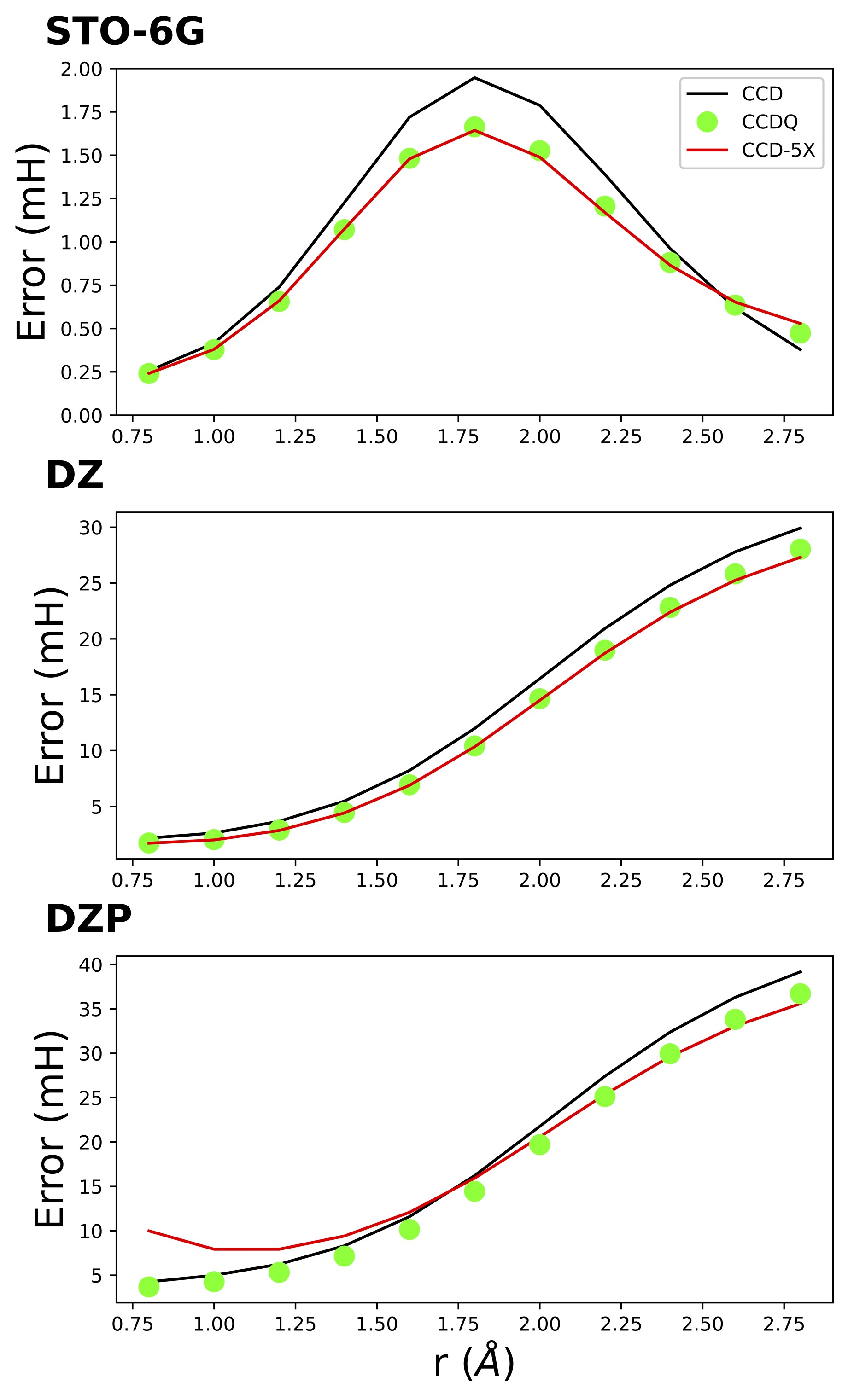}
\caption{Error analysis of the single OH bond dissociation of H$_2$O, using the STO-6G, DZ and DZP basis sets. Internal HOH angle is constrained to 104.5 degrees. Error is w.r.t FCI for STO-6G/DZ basis sets, but CCSDTQPH for the DZP basis.  }
\label{fig:H2Osingle}
\end{figure}

To ascertain the utility of augmented $T_2$-based approaches, we initially study the single OH-bond dissociation of the water molecule. As it was found that the differences separating the various orders of PT-CCD($k$), CCD-$k$X, and XCCD($k$) were minimal (routinely within 1 mH), Figure \ref{fig:H2Osingle} only compares the error of CCDQ and the iterative CCD-5X method against FCI/CCSDTQPH in various basis sets. This fact indicates that higher-order, $T_6$-like operators are not likely to be necessary in weakly-correlated examples. In fact, even adding the lowest, fifth-order diagram to the CCD-5X residual equations only marginally improves upon the error of baseline CCD  in the STO-6G and DZ basis sets. However, the error for CCD-5X does exceed that of CCD in the bound region of H$_2$O for the DZP basis set, which is a trend that nevertheless reverses upon dissociation. Regardless, the CCD-5X method generally tracks closer to CCDQ results than the baseline CCD method.


\clearpage

\subsection{N$_2$}

\begin{figure}
\includegraphics[scale=0.125]{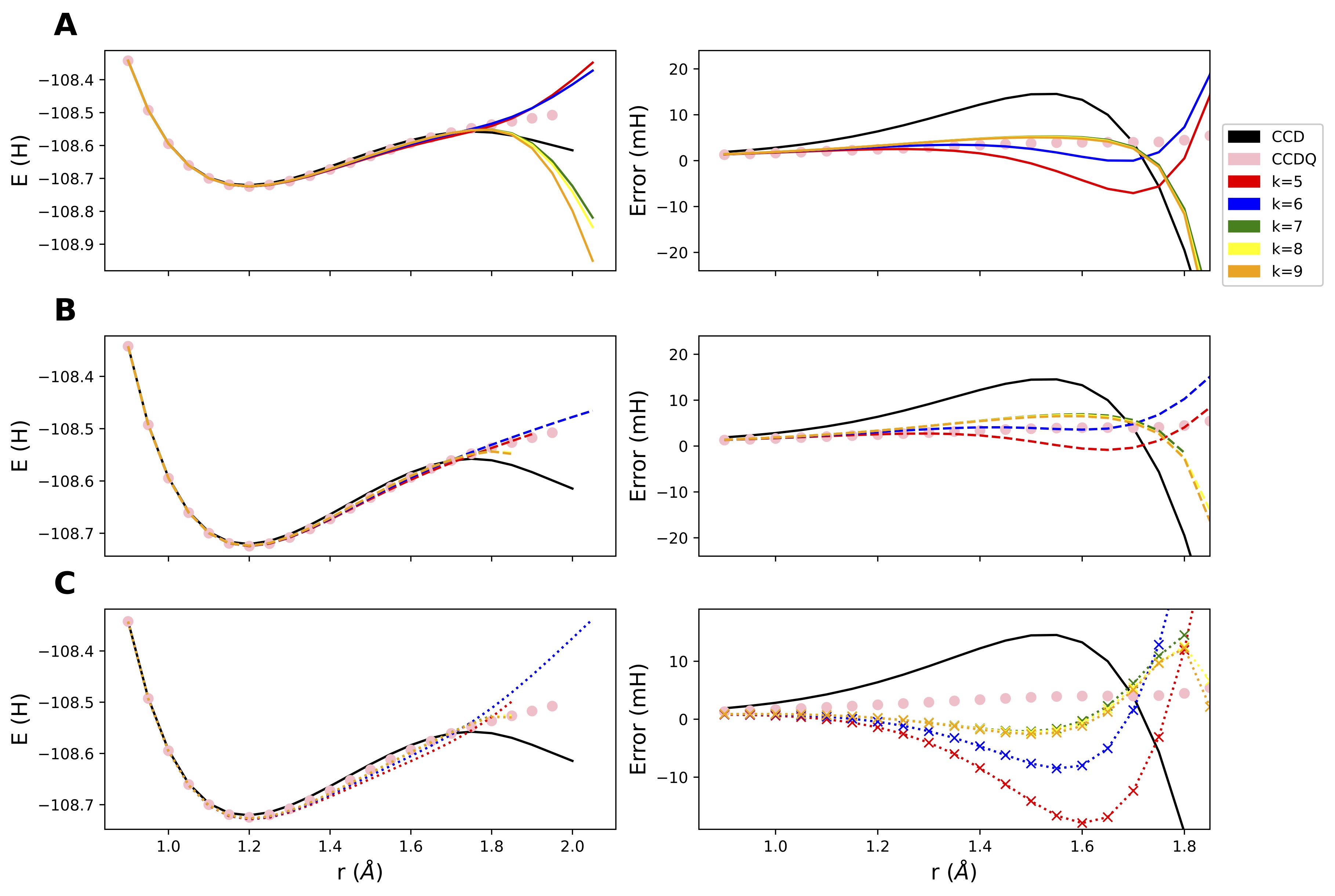}
\caption{Surface and error comparison of select CC/XCC-based methods with respect to FCI for the dissociation of N$_2$ in the STO-6G basis set. The row of plots in A) illustrate the PT-CCD($k$) results, B) depict CCD-$k$X results, and C) show the XCCD($k$) results. Note the legend defines order $k$ based on a color scheme. }
\label{fig:N2bond}
\end{figure}

The performance of the augmented $T_2$ methods becomes even more distinct in the dissociation of N$_2$. Figure \ref{fig:N2bond} analyzes this process with respect to the STO-6G basis set. The purely perturbative PT-CCD($k$) approximations displayed in row A exhibit better agreement with FCI up to 1.8 $\AA $ as compared to standard CCD. Beyond this point, the poor quality of the reference determinant leads to divergent behavior where we see that purely $T_4$-like orders ($k=5,6$) tend to diverge toward $+\infty$ while orders that  include $T_6$-like operators ($k=7,8,9$) tend to diverge toward $-\infty$. Because this oscillatory behavior is predictable, it can be compensated for by performing either a high-order polynomial fit, or some other non-linear fit like Pade approximant, which is well-known to provide converged MBPT results for apparent divergent approximations. This technique has been used before,\cite{hirata2000high} but is not pursued in the context of the current work. 

Regarding the purely iterative CCD-$k$X class of methods shown in Figure \ref{fig:N2bond} B, we note that the error of all methods tend to be superior than those exhibited by standard CCD when the residual equations could be converged. As can be seen, it is no small feat to converge the residual equations even when varying DIIS parameters. This phenomena similarly extends into the expectation-value CCD methods of Figure \ref{fig:N2bond} C. Curiously, it seems adding the extra sixth-order, $T_4$-like diagram tends to aid convergence of the underlying residual equations as shown by the CCD-6X and XCCD(6) methods. Figure \ref{fig:N2bond} C also shows the first evidence that higher-order $k=7,8,9$ terms net any tangible gain over their lower-order $k=5,6$ counterparts, although inclusion of these diagrams to the residual equations significantly degrades convergence behavior. Nevertheless, it appears that little is to be gained by adding in higher-order $T_6$-like diagrams beyond the lowest-order ($n=7$) for this example. 

Note that the STO-6G basis of N$_2$ allots 6 virtual spin-orbitals and therefore would support up to hextuple excitations out the reference space. Nevertheless, the CCDQ method which explicitly omits any description of the $T_1$, $T_3$, $T_5$, and $T_6$ tensors exhibits negligible error across the PES as compared to the FCI. This seems to indicate the importance of added quadruple excitations on top of CCD, which the residual equations of XCCD-based methods implicitly support. Of course, we provide CCDQ results where the equations could be converged. It is expected that even full quadruples is not sufficient beyond ~2 \AA.

\begin{figure}
\includegraphics[scale=0.25]{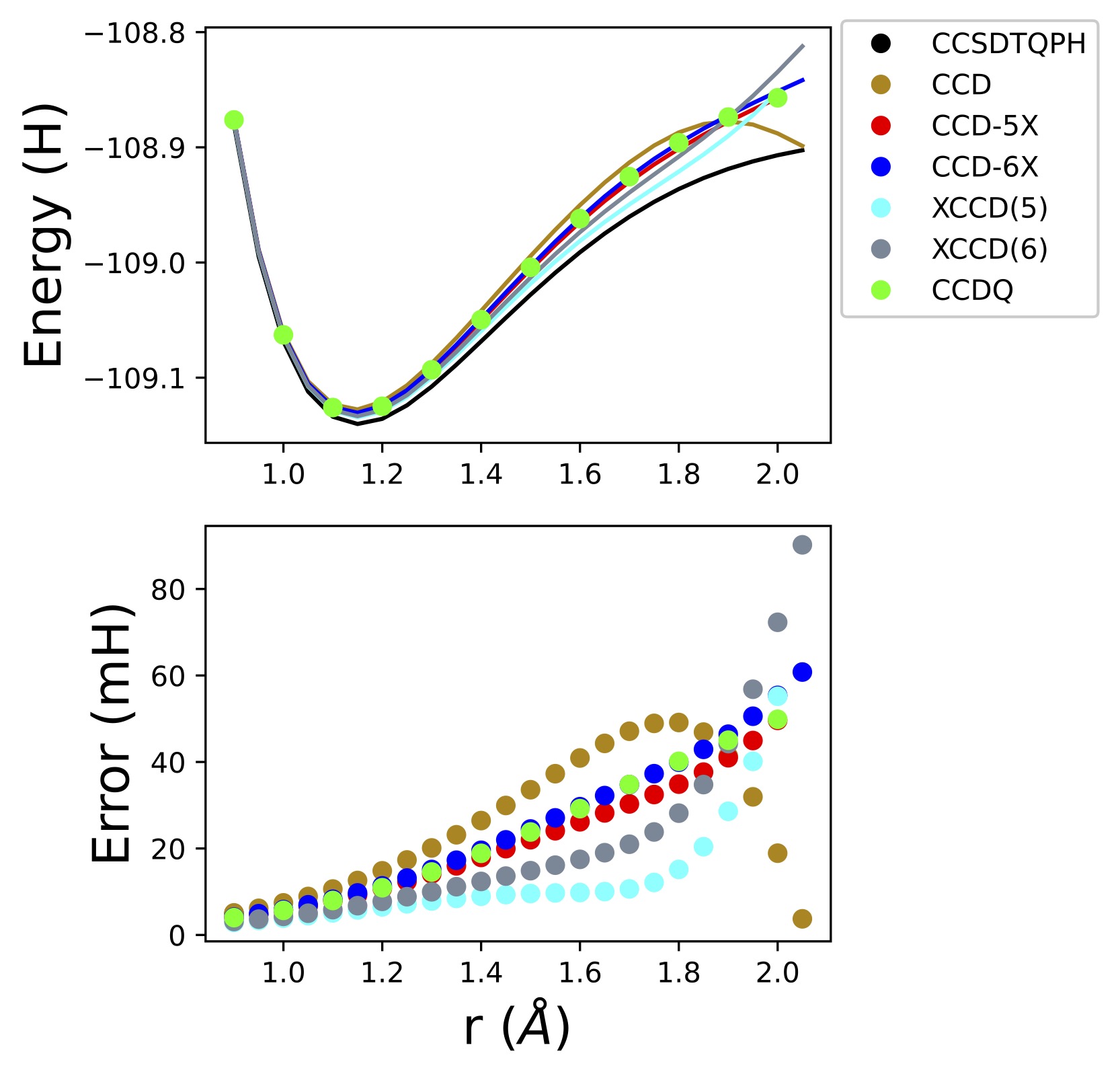}
\caption{Surface and error comparison of select CC/XCC-based methods with respect to CCSDTQPH for the dissociation of N$_2$ in the DZ basis set. }
\label{fig:N2dz}
\end{figure}
Figure \ref{fig:N2dz} exhibits the N$_2$ PES and associated errors of the fifth and sixth order methods in the DZ basis set. In either case the XCCD($k$) methods show better agreement with the ``exact" answer at shorter bond distances, but the added perturbative corrections to the CCD energy (as shown by the $k=5,6$ diagrams in Figure \ref{fig:XCCDenergy}) ultimately lead to erroneous behavior near the dissociated limit. We know these energy diagrams are the culprits behind this behavior in XCCD($k$)  because analogous behavior is \textbf{not} emulated in the corresponding set of CCD-$k$X methods, which both track closely to CCDQ results. Nevertheless, we again find that the convergence of the residual equations are more stable when the sixth-order term is added to the residual equations.

\clearpage

\clearpage

\subsection{H$_4$ ring}

\begin{figure}
\includegraphics[scale=0.25]{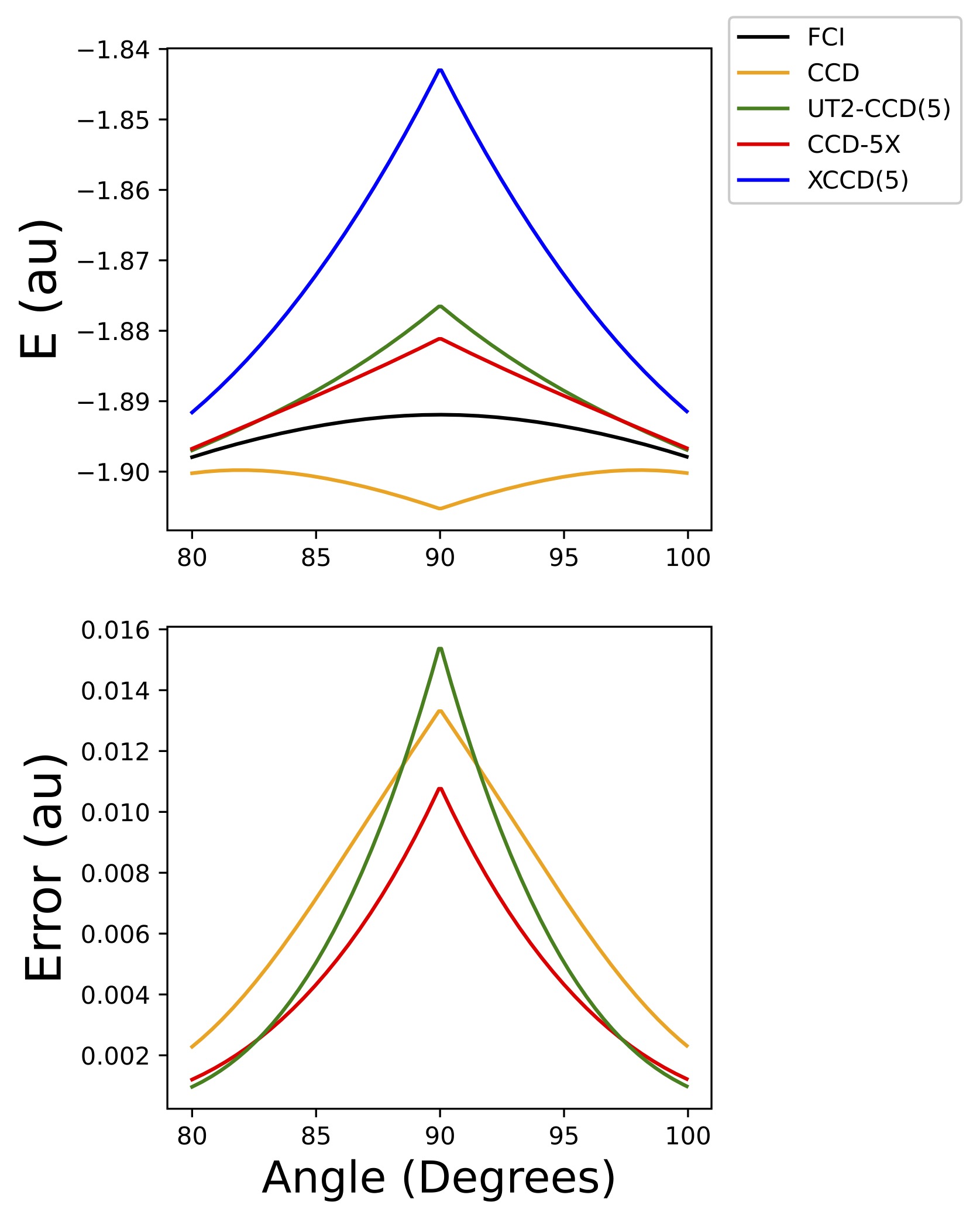}
\caption{H$_4$ ring scanned over the internal rotation angle }
\label{fig:H4ring}
\end{figure}

When the $H_4$ ring example is studied,  clear benefits are shown by the family of augmented $T_2$ methods. In this case, it has been established that traditional CC methods struggle for this deceivingly simple example.\cite{paldus1993application,piecuch1994application} For the current work, the H atoms are positioned around the circumference of a circle which has a diameter of 6.569 Bohr. We vary the angle subtended by lines connecting opposing H atoms between 80 and 100 degrees, thereby mimicking the geometries studied recently.\cite{marie2021variational} Figure \ref{fig:H4ring} shows the fifth-order perturbative, purely iterative, and XCCD methods for the ring rotation of H$_4$ in the STO-6G basis set. In this basis set, fifth-order methods were the only members of this hierarchy that had meaningful contributions to the correlation energy. 

In this example, the first notable trait of these methods is that they contribute a net positive correction to the correlation energy. As a consequence, each method overshoots the FCI, which is in stark contrast to standard CCD. Furthermore, each of these methods largely follow the variational CCD (VCCD) trend\cite{van2000benchmark} despite not rigorously satisfying an upper bound to FCI. We find that in regions outside $\pm$ 3 degrees from the critical point found at 90 degrees, there is clear improvement over CCD in both the perturbative PT-CCD(5) and iterative CCD-5X approximation. However, we note that while the perturbative approximation exhibits errors that exceed CCD at $\pm$ 2 degrees around the critical point, the iterative CCD-5X approximation reliably outperforms CCD over the entire interval. The error of expectation value CCD (XCCD(5)) is not plotted, as it exceeds that of standard CCD across the interval and skews the plot. 

\begin{figure}
\includegraphics[scale=0.15]{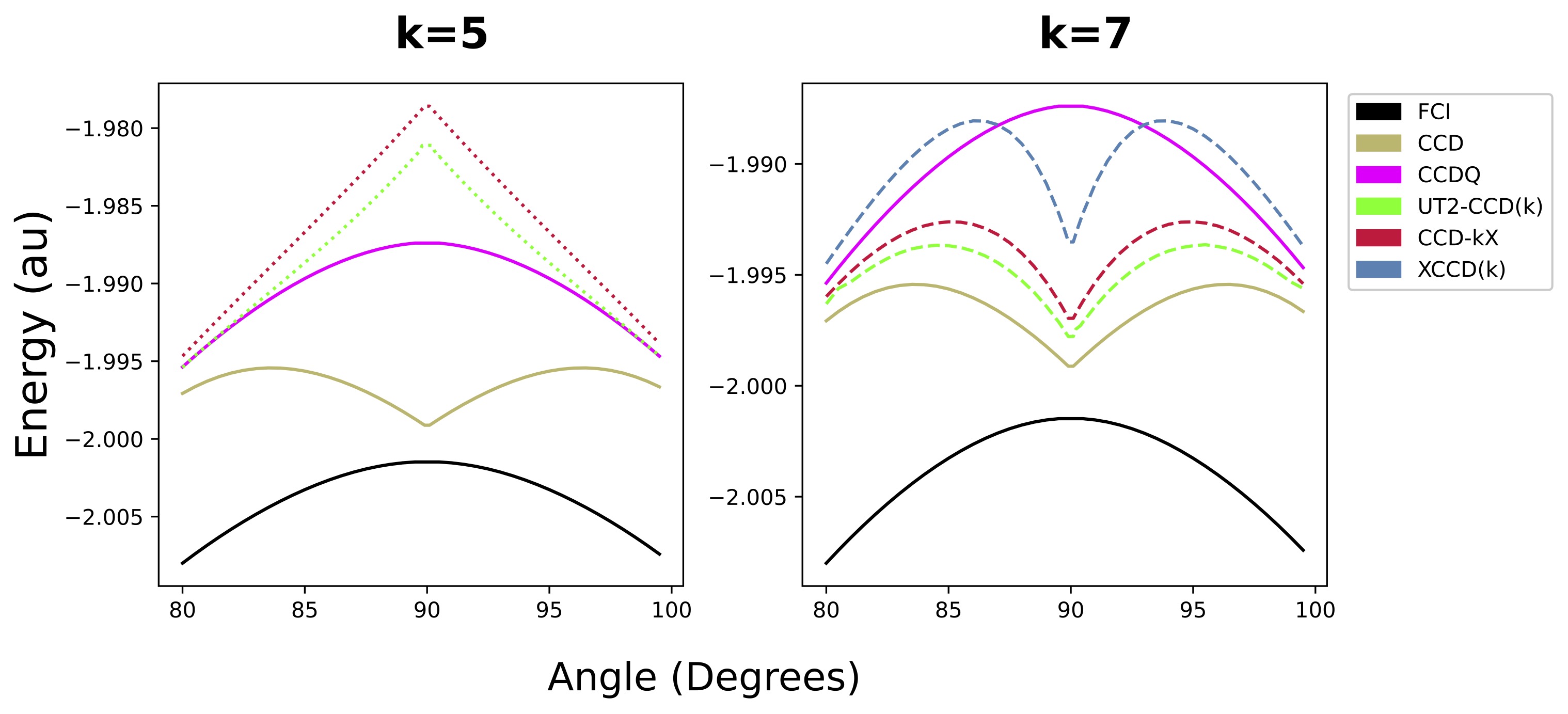}
\caption{Surface generated by select FCI and CC/XCC-based methods using the DZ basis for the H$_4$ ring. A) illustrates the $k=5$ results, whereas B) illustrates the $k=7$ results. }
\label{fig:H4DZbasis}
\end{figure}

Figure \ref{fig:H4DZbasis} showcases the same H$_4$ example in the DZ basis using methods that are complete through orders $k=5$ and $k=7$. In this example, inclusion of the leading-order $k=7$ diagram in the current work's hierarchy of methods yield meaningful contributions to the overall correlation energy. However, it was found that incorporating diagrams associated with higher-order, $T_4$ and $T_6$-like excitations do not provide large impacts to the correlation energy. 

For this basis set, we are able to converge the CCDQ equations and include the results as reference values. The CCDQ method exhibits analogous behavior as the FCI, albeit shifted by $\sim$10 mH which is attributable to the missed single, triple, etc excitations in the underlying method. On the other hand, the CCD method yields a PES with the incorrect curvature that is energetically bound by CCDQ and FCI. Clearly, the inclusion of explicit quadruples is important in this example/basis set. 

The $k=5$ ordered methods overshoot the CCDQ results and exhibit the same critical point at 90 degrees that was seen in the STO-6G basis. Note that the XCCD(5) method is not shown here, as it behaves similarly to the STO-6G example and would skew the plot. At fifth-order we see that the PT-CCD(5) method actually shows closest agreement to CCDQ, but even the iterative CCD-5X method agrees to $\sim$10 mH at the critical point. 

The PES behavior dramatically changes once the lowest-order $T_6$-like excitations are added to our hierarchy of methods. As shown by the $k=7$ results of Figure \ref{fig:H4DZbasis}, the overall curvature of the PES emulates the incorrect behavior of CCD yet is energetically closer to CCDQ. At this order, the method in most agreement with CCDQ is now XCCD(7), which is in error by no more than $\sim$5 mH.

\clearpage

\subsection{C$_2$H$_2$ torsion}

The torsional rotation of ethylene  undergoes a diradical transition at a critical point of 90 degrees, wherein the $\pi$ and $\pi^{*}$ molecular orbitals become degenerate. A single, doubly-occupied Slater is not equipped to resolve the inherent multi-reference behavior in such a situation, where the ``correct" zeroth order description might include equally weighted consideration of the $(\pi)^2$ and $(\pi^*)^2$ electronic configurations. The lack of flexibility to properly account for this second configuration within a RHF reference ultimately leads to a unphysical cusp at 90 degrees, which is also found in low-rank CC methods (e.g. CCSD).\cite{krylov1998size} This system, and others like it, have been studied extensively\cite{wood1974barrier,staemmler1977note,krylov2002perturbative,hoffmann2009comparative,musial2011multi,musial2011multireference} with the general consensus being unmodified, low-rank CC is not suitable to describe this PES and that alternative approaches must be sought.\cite{lyakh2011tailored,melnichuk2014relaxed}

\begin{figure}
\includegraphics[scale=0.25]{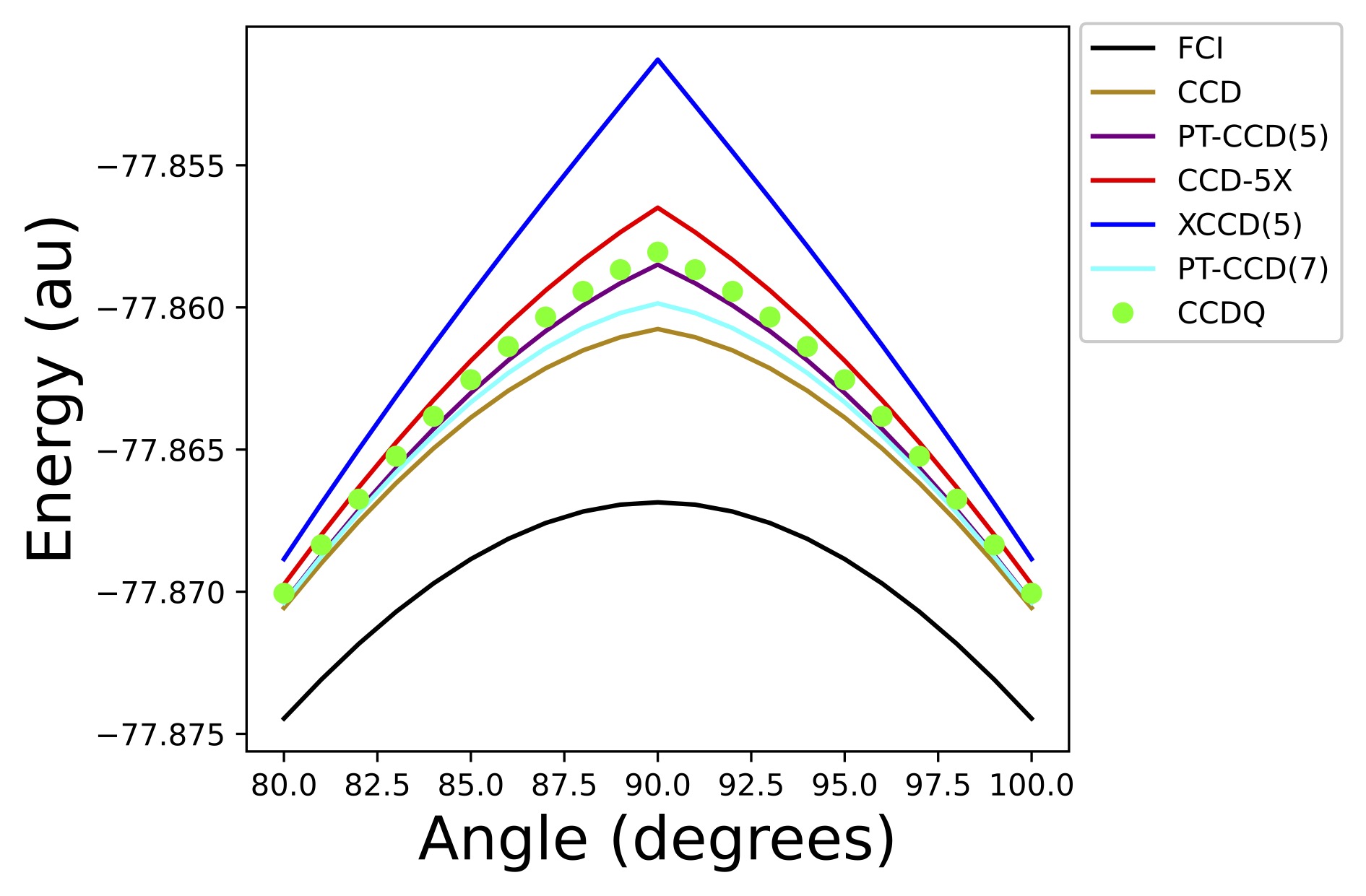}
\caption{Minimal STO-6G basis scan of the internal rotation angle of ethlyene using an assortment of methods.}
\label{fig:ethSTO6G}
\end{figure}

Figure \ref{fig:ethSTO6G} illustrates an assortment of methods' performance for this process in the STO-6G basis set. All methods shown exhibit a slight cusp at 90 degrees, although to varying extends. Curiously, the standard CCD method is actually in the closest agreement with FCI in this basis set, surpassing the more ``complete" CCDQ method by $\sim$5 mH at the critical point. We find that while CCD-5X and PT-CCD(7) bound the CCDQ surface by roughhly equal and opposite magnitudes, the purely perturbative PT-CCD(5) method actually exhibits the closest agreement with CCDQ. Just as in the STO-6G basis set example of H$_4$, the XCCD(5) results largely overshoot the FCI result and has both the sharpest cusp and largest error.

\begin{figure}
\includegraphics[scale=0.25]{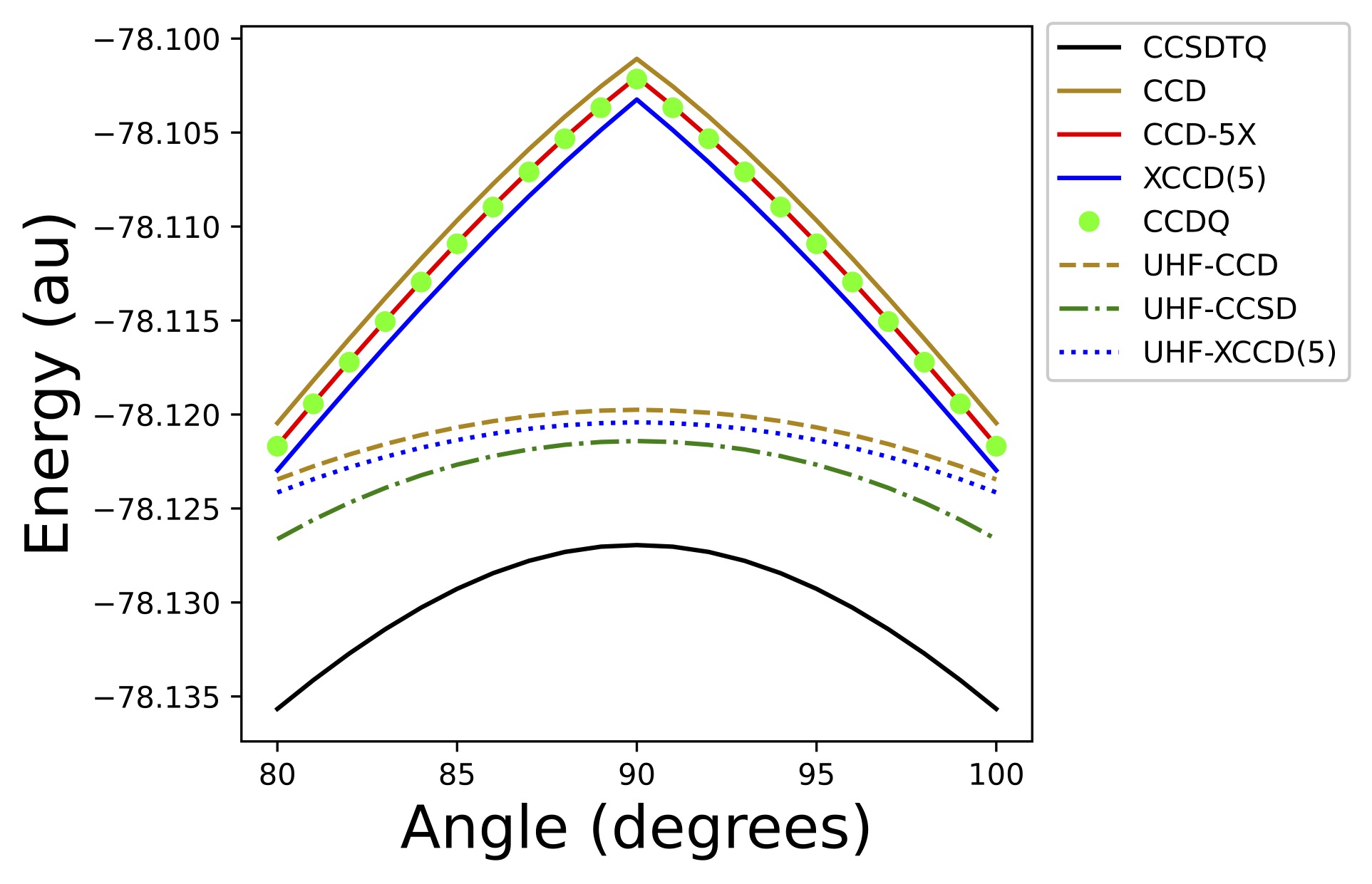}
\caption{DZ basis scan of the internal rotation angle of ethlyene. RHF reference unless otherwise specified. }
\label{fig:ethDZ}
\end{figure}

After increasing the basis set to DZ quality, our standard for comparison becomes RHF-CCSDTQ as this was what could be afforded in the allotted timeframe. Figure \ref{fig:ethDZ} compares all fifth-order methods against CCD and CCDQ, where we note that minimal improvements were found going up to higher orders in this basis set. In this example, the trend found in the STO-6G case flips wherein the baseline CCD method incurs the largest error while XCCD(5) shows the best overall agreement with CCSDTQ. Notably, the spread between these methods' predictions is consistently between $\sim$1-2 mH. While this may be the case, the CCD-5X method is actually in closest agreement with CCDQ. Regardless, all methods are in error with CCSDTQ by $\sim$25 mH in the worst case - the critical point - and are not capable of eliminating the spurious cusp. 

We also provide results of broken symmetry-based calculations as well, with UHF-CCSD added for comparison purposes. It should be noted that while using a UHF reference rectifies the spurious cusp, this comes at the expense of a significant degradation in spin quantum numbers. We found that UHF-based CCD-5X and XCCD(5) both slightly improve upon UHF-CCD across in the PES with the same energetic ordering as in the RHF-based calculations.

\clearpage

\section{Conclusion}

In this work, we present a hierarchy of methods that can be constructed from considerations of finite-order, expectation-value CC theory. This hierarchy is in turn  composed of three classes of methods: one includes purely perturbative corrections to the CCD energy, another seeks to  augment the $T_2$ residual equations of CCD by additional terms, and the last is a hybrid approach which combines both of the aforementioned approaches simultaneously and represents finite-order, XCCD theory. Our primary objective is to explore the efficacy and limitations of this hierarchy of methods, in our pursuit of what we call an ``ultimate" $T_2$ method. To this end, a tangential goal of the current work is to efficiently emulate the behavior of higher-rank CC methods that explicitly account for even-ordered electron excitations - like CCDQ - without paying the associated ($\mathcal{O}(N^{10})$) cost. As has already been discussed,\cite{strongCorrT2} our current approach based on finite-orders of XCCD naturally lends itself to efficient, factorized intermediate schemes that nevertheless implicitly account for higher-order excitations.  Notably, we find that lowest-order breed of methods, like CCD-5X, are routinely in close agreement with CCDQ but nevertheless scale like CCD: $\mathcal{O}(N^6)$. Furthermore, in situations emblematic of ``strong" electron correlation we find that the proposed hierarchy of methods rather consistently improve upon baseline CCD. Similar conclusions were drawn in prior work studying the 6 and 10-site, half-filled Hubbard Hamiltonian.\cite{strongCorrT2}

As a final consideration, it is prudent to mention the concerns brought up by Szalay et al.\cite{szalay1995alternative} regarding the full XCC method that does not truncate the cluster operator. The point is that XCC(5) forces the cancellation of terms associated with $T_4$ in the energy functional that incurs an error of $\delta(6)$, noting the absence of a term like $\braket{0|T_4^{\dagger}f_NT_4|0}$ which first shows up at sixth order. As a byproduct of this incomplete cancellation, the corresponding residual equations incur an error $\delta(4)$. This means that the corresponding wavefunction can only be correct through third order. Although this point does not necessarily impact the validity of the current work focusing exclusively on $T_2$, a new approach under the XCC framework that properly accounts for both the energy functional \textbf{and} the wave function are called for. This will be further explored in follow-up work.

\section*{Acknowledgements}
This work was supported by the Air Force Office of Scientific Research under AFOSR Award No. FA9550-23-1-0118. Z.W.W. thanks the National Science Foundation and the Molecular Sciences Software Institute for financial support under Grant No. CHE-2136142. Z.W.W. also acknowledges support from the U.S. Department of Energy, Office of Science, Office of Workforce Development for Teachers and Scientists, Office of Science Graduate Student Research (SCGSR) program. The SCGSR program is administered by the Oak Ridge Institute for Science and Education (ORISE) for the DOE. ORISE is managed by ORAU under contract number DE-SC0014664.

\section{Appendix}

\subsection{XCCD expressions at arbitrary orders}
We note that the tediousness of the prior derivations originate from the desire to cancel internally disconnected diagrams in the energy functional. Our strategy is based on our knowledge of what happens to these terms in the unterminated limit. If  the energy functional $\braket{0|e^{T_2^{\dagger}}H_Ne^{T_2}|0}_c$ is not  terminated prematurely, these internally disconnected diagrams will ultimately cancel. So our strategy in some sense seeks to short-circuit a path to this limit even in finite-orders. To generalize our strategy above, we could simply enforce the condition that internally disconnected terms are 0 from the beginning, meaning that only the connected portion of the energy arising at odd orders is allowed. The same is true for the only connected term that would show up at even orders as well. But we are looking for a more general proof.  

We note that by analyzing the the above developments, some patterns begin to emerge. It is helpful to start by isolating the terms, order-by-order, in the generating functional that contained internally disconnected pieces. Starting with the even orders $p=2n$, we find that  terms like $\braket{0|\bigg(\frac{T_2^{\dagger}}{n!}\bigg)^n\bigg(W_nT_2^{n-1} + f_NT_2^n \bigg)|0}$ can always be immediately omitted from the generating energy functional since 
\begin{equation}
    \begin{split}
        \bigg( \frac{\big(T_2^{\dagger}\big)^{n}}{n!}\Bigg[ Q_2(W_N + f_NT_2)\Bigg]\frac{T_2^{n-1}}{(n-1)!}\bigg)=0
    \end{split}
\end{equation} where the term in the square bracket are the XCCD(2) equations, meaning the  error associated with this cancellation would be $\delta(2+n+(n-1))$.  We note that similar strategies can be pursued to eliminate this genre of terms at arbitrary even orders. 


Consequently, we are only left with internally connected terms of the form $\braket{\phi^{n-1}|W|\phi^n}$, which can contribute to the energy. The internally disconnected portion - which requires cancellation - is of the form
\begin{equation}\label{eq:evenOrders}
\braket{0|\sum_{n=2}^{\infty}\frac{(T_2^{\dagger})^x}{(n-1)!}\bigg( (T_2^{\dagger})^{\lfloor\frac{n-2}{3}\rfloor}W_N\frac{T_2^n}{n!}\bigg)_D|0}
\end{equation} for $x=(n-1) - \lfloor \frac{n-2}{3} \rfloor$.

Turning to the odd order contributions to the energy functional, we find that the internally disconnected portions can be found by the generating sequence
\begin{equation}\label{eq:oddOrders}
    \braket{0|\sum_{n=2}^{\infty}\frac{(T_2^{\dagger})^y}{n!}\bigg( (T_2^{\dagger})^{\lfloor \frac{n}{3}\rfloor}\frac{W_NT_2^n}{n!}\bigg)_D|0}
\end{equation} for $y=n-\lfloor \frac{n}{3}\rfloor$

Noting that the (connected) residual equations take the general form 
\begin{equation}\label{eq:generalResidEqns}
    Q_2\bigg( f_N + W_N + W_NT_2 + \sum_{n=2}^{\frac{k-1}{2}-1}\frac{(T_2^{\dagger})^{n-2}}{(n-2)!}\frac{W_NT_2^n}{n!} + \sum_{n=2}^{\frac{k}{2}-1}\frac{(T_2^{\dagger})^{n-1}}{(n-1)!}\frac{W_NT_2^n}{n!}\bigg)_c=0
\end{equation} for order $p$, the order $k=p-2$ residual equations can be invoked to explicitly canceled the internally disconnected contributions arising from Equations \eqref{eq:evenOrders} and \eqref{eq:oddOrders} once a factor of $\frac{(T_2^{\dagger})^2T_2}{2}$ have been pulled out of either. This process leaves only internally connected diagrams in the energy functional. 

The base cases for this are orders $p=5$ and $p=6$, wherein we have already shown that the orders $p-2$ residual equations cancel the disconnected contributions to the energy functional in either case. The induction step assumes the same is true for arbitrary even and odd orders. Thus, we need to show that this is true at orders $p$ and $p+1$, for the arbitrary odd and even situation respectively. We choose to work with orders characterized by the sequence $\{ p_0-2,p_0-1,p_0,p_0+1,p_0+2,p_0+3\}$ to prove our point. Let's start with the highest order being even, and of the form $2n=p_0+1$. We want to show that the resulting energy functional contains the $2n=p_0-1$ residual equations that are cancel the internally disconnected pieces. 
We find it easier to proceed from the original XCCD energy functional wherein we find that up to arbitrary even order
\begin{equation}
\begin{split}
    \Delta E &= \cdots \braket{\phi^{\frac{p-1}{2}}|W_N|\phi^{\frac{p-1}{2}}} + \braket{\phi^{\frac{p-1}{2}}|W_N|\phi^{\frac{p+1}{2}}} + \braket{\phi^{\frac{p+1}{2}}|W_N|\phi^{\frac{p+1}{2}}} 
\end{split}
\end{equation} which - after factoring out  $(T_2^{\dagger})^{2}T_2$ - can be simplified such that
\begin{equation}\label{eq:evenFullFunc}
    \begin{split}
        \Delta E &= (T_2^{\dagger})^2T_2\bigg( f_NT_2+W_N+W_NT_2\cdots+ (T_2^{\dagger})^{\frac{p-5}{2}}W_NT_2^{\frac{p-3}{2}} \\
        &+ (T_2^{\dagger})^{\frac{p-5}{2}}W_NT_2^{\frac{p-1}{2}} + (T_2^{\dagger})^{\frac{p-3}{2}}W_NT_2^{\frac{p-1}{2}}\bigg)
    \end{split}
\end{equation} Next, we need to expand Equation \eqref{eq:generalResidEqns} so that the maximal term of the resulting residual equation is of (even) order $2n=p_0-1$. Doing this, we see that 
\begin{equation}\label{eq:evenresidual}
\begin{split}
    Q_2\bigg(& f_NT_2+W_N + W_NT_2+\cdots+ (T_2^{\dagger})^{\frac{p-3}{2}-1}W_NT_2^{\frac{p-3}{2}}\\
    &+ (T_2^{\dagger})^{\frac{p-1}{2}-2}W_NT_2^{\frac{p-1}{2}} +(T_2^{\dagger})^{\frac{p-1}{2}-1}W_NT_2^{\frac{p-1}{2}}\bigg)=0
\end{split}
\end{equation} Thus, we have shown that the internally disconnected pieces generated by Equation \eqref{eq:evenFullFunc} at order $p_0+1$ can be eliminated by invoking the residual equations at order $p_0-1$.

The same logic extends to arbitrary odd orders, where we can add the next odd term in the sequence found by $2n+1=p_0+3$ to Equation \eqref{eq:evenFullFunc} -given by $\braket{\phi^{\frac{p+1}{2}}|W_N|\phi^{\frac{p+3}{2}}}$- and after factoring out $(T_2^{\dagger})^{2}T_2$ from the resulting energy functional, find that the result is Equation \eqref{eq:evenresidual} plus $(T_2^{\dagger})^{\frac{p-3}{2}}W_NT_2^{\frac{p+1}{2}}$, or equivalently, the (odd) order $2n+1=p_0+1$ residual equations.

After these cancellation of internally disconnected terms has been assured, it becomes evident that starting at fourth order, only one term - which is internally connected - is allowed to contribute to the energy at each successive order. This is similarly the case for the residual equations, which is alluded to by Equation \eqref{eq:generalResidEqns}.

\subsection{Factorization theorem for arbitrary $T_{2n}$}
We now seek to generalize the above formulations in order to systematically generate improved working $T_2$ equations based off a cluster operator truncated at arbitrarily high-order $2n$, equivalently, $T_{2n}$. In order to provide further detail on such an ``ultimate", $T_2$-based CC method, the following induction proof succinctly summarizes how many of the tedious details covered in the preceding sections can be avoided by first recognizing the recurrence relation present between sequential orders, and then take advantage of the factorization pattern.

To facilitate book-keeping, we prefer to associate the $n$ available options for time-ordering that any single $W_n$ has using characters in a set. Such a set, defined as $\mathcal{X}$, contains $n$ unique characters, $\{ A, B, C, \cdots , N\}$. A power set can then be generated from this alphabet, $\mathcal{P}(\mathcal{X})$, from which one of the resulting $2^N$ subsets are assigned as a superscript to a denominator. Note that for $r \in [n,n-1,\cdots,1]$ defining a general cluster order, $2r$, and its denominator, there are $C(n,r)=\frac{n!}{(n-r)!r!}$ possible combinations of $\mathcal{P}(\mathcal{X})$ available for assignment to a denominator, where each contains a string of $\frac{n!}{(n-r)!r!}$ elements from the alphabet, $\mathcal{X}$. Every denominator follows this character assignment. So, the solitary $D_{2n}$ denominator is assigned the entire set $\mathcal{X}$ as a superscript, the $C(n,n-1)=\frac{n!}{(n-1)!}$ distinct $D_{2(n-1)}$ denominators are each uniquely assigned to one of the $C(n,n-1)=\frac{n!}{(n-1)!}$ possible subset combinations of  $\mathcal{P}(\mathcal{X})$ that contain as many characters, the $C(n,n-2)=\frac{n!}{(2)!(n-2)!}$ distinct $D_{2(n-2)}$ denominators are each uniquely assigned to one of the $C(n,n-2)=\frac{n!}{(2)!(n-2)!}$ possible combinations of   $\mathcal{P}(\mathcal{X})$ that also contain as many characters, continuing in this way until $r=1$.

With these details in mind, we proceed with the induction proof where we want to prove the following proposition thru order $n$ subject to $P(-1)=0$:
\begin{equation}\label{eqn:IndProp}
P(n, \mathcal{Z}):  \frac{1}{n!}\frac{1}{D_2^AD_2^BD_2^C\cdots D_2^N} = \frac{1}{D_{2N}^{ABC\cdots MN}} \sum_{y\in \mathcal{Z}}P\bigg( n-1,  y \bigg)  
\end{equation} In this case, the set $\mathcal{Z}$ further consists of $C(n,n-1)$ subsets, wherein each contains as many character combinations. The subsets of  $\mathcal{Z}$ are summed over to ensure the recursively-called proposition returns results that exhaustively considers all available time-orderings. In this case, the LHS of Equation \ref{eqn:IndProp} is the product of $D_2$ denominators at various time-orders and the RHS expresses the product between  the (massively-dimensioned) $4n$-index denominator and  recursions of the proposition that will cycle through all lower-orders.

We will include $n=1, 2$ as base cases to be clear. Inserting the $n=1$ case into Equation \ref{eqn:IndProp}, we find that $P(1,\{A\}): \frac{1}{D_2^A} = \frac{1}{D_2^A}$ as expected. For $n=2$, we find agreement that 
\begin{equation}
\begin{split}
    P(2,\{ \{A\}, \{B\}\}): \frac{1}{2}\frac{1}{D_2^AD_2^B} &= \frac{1}{D_4^{AB}}\bigg( P(1,A) + P(1, B) \bigg) \\
    &= \frac{1}{D_4^{AB}}\bigg( \frac{1}{D_2^A} + \frac{1}{D_2^B}\bigg)
    \end{split}
\end{equation} which coincides with the results of earlier subsections focusing on $T_4$ contributions to $T_2$ by virtue of factorizing the $D_4$ denominator. 

For the inductive step, we assert that the proposition in Equation \ref{eqn:IndProp} is true. We now wish to prove the proposition $P(n+1, \mathcal{Z} + O)$ such that
\begin{equation}\label{eqn:finalIndcStep}
\frac{1}{(n+1)!}\frac{1}{D_2^AD_2^BD_2^C\cdots D_2^ND_2^O} = \frac{1}{D_{2(N+1)}^{ABC\cdots MNO}} \sum_{y\in \mathcal{A}} P\bigg( n+1,  y \bigg)
\end{equation} where $\mathcal{A}$ now considers the extra character $O$ alongside the existing alphabet when generating the $C(n+1,n)$ subset combinations. By rearranging the newly-updated Equation \ref{eqn:IndProp} for $\sum_{y \in \mathcal{A}} P(n+1,y)$, the resulting expression can be inserted back into the RHS of Equation \ref{eqn:finalIndcStep}, leading to the final result
\begin{equation}
\frac{D_{2(N+1)}^{ABC\cdots MNO}}{\bigg(D_{2(N+1)}^{ABC\cdots MNO}\bigg) \bigg(D_2^AD_2^BD_2^C\cdots D_2^ND_2^O\bigg)} = \frac{1}{D_2^AD_2^BD_2^C\cdots D_2^ND_2^O} 
\end{equation} which recovers the LHS of Equation \ref{eqn:finalIndcStep}, successfully concluding our inductive proof. 

This definitively establishes that the procedure of using arbitrarily high-order cluster operator, $T_{2n}$, to sequentially and systematically improve the fidelity of $T_2$ comes with the benefit of eliminating all (expensive) $>4$-index denominators from the resulting equations, in exchange for $n$ products of $4$-index denominators. Further opportunities for profound savings in computational expense are also indirectly supported, since such a structure more easily supports the construction of intermediates. 

\subsection{CC functional route toward ultimate $T_2$ method}

In order to strictly abide by - and invoke - the factorization theorem as used in the context of MBPT, we find it useful to work with  the CC functional instead of Equation \ref{eq:corrE}
\begin{equation}\label{eq:CCfunctional}
E_{CC}=\braket{0|(1+\Lambda)(H_Ne^{T})_c|0}
\end{equation} where $\Lambda$ serves as a deexcitation operator having the general form
\begin{equation}
\begin{split}
    \Lambda =& \bigg(\frac{1}{n!}\bigg)^2\sum_{ijkabc\cdots}\lambda_{abc\cdots}^{ijk\cdots}\{i^{\dag}aj^{\dag}bk^{\dag}\cdots \} \\
    =& \Lambda_1 + \Lambda_2 +\Lambda_3 + \cdots
    \end{split}
\end{equation} and which acts on $1, 2, 3, \cdots$ particle/hole pairs. To correspond with the theme of this work, a $\Lambda$-based ultimate $T_{2n}$ theory will have
\begin{equation}
\begin{split}
        \Lambda(k\rightarrow \infty)&= \sum_{n=1}^{k} \frac{\Lambda_{2n}^{(n)}}{n!}\\
        T&=\sum_{n=1}^k T_{2n}
\end{split}    
\end{equation}

However, in the current work we truncate up to hexatuple excitations by making the following approximations
\begin{equation}
\begin{split}
    \Lambda = &\Lambda_2^{[1]} + \frac{\Lambda_4^{[2]}}{2} + \frac{\Lambda_6^{[3]}}{3!} \\
    T =& T_2 + T_4^{[3,4]} + T_6^{[5,6]}
\end{split}
\end{equation} 
where $\Lambda_{2n}$ is expressed in terms of the adjoint of a first-order $T_2$ given by 
\begin{equation}
    \Lambda_{2n}^{(n)}=\prod_{i=1}^n T_2^{(1)*}
\end{equation} for $T_2^{(1)*} = \frac{\braket{ij||ab}}{D_2}$. 
The superscripts refer to the order(s) in which a diagram is correct through in PT. This particular choice in approximation is made because choosing $\Lambda_{2n}^{(n)}$ to be proportional to $n$ products of $T_2^{(1)*}$ directly facilitates future use of the factorization theorem to cancel higher rank denominators. This further offers an additional opportunity to significantly cheapen the resulting calculation by avoiding explicit solution of the infinite-order $\Lambda$ equations; hopefully, without significantly compromising the resulting method's integrity.


\begin{figure}
\includegraphics[scale=0.25]{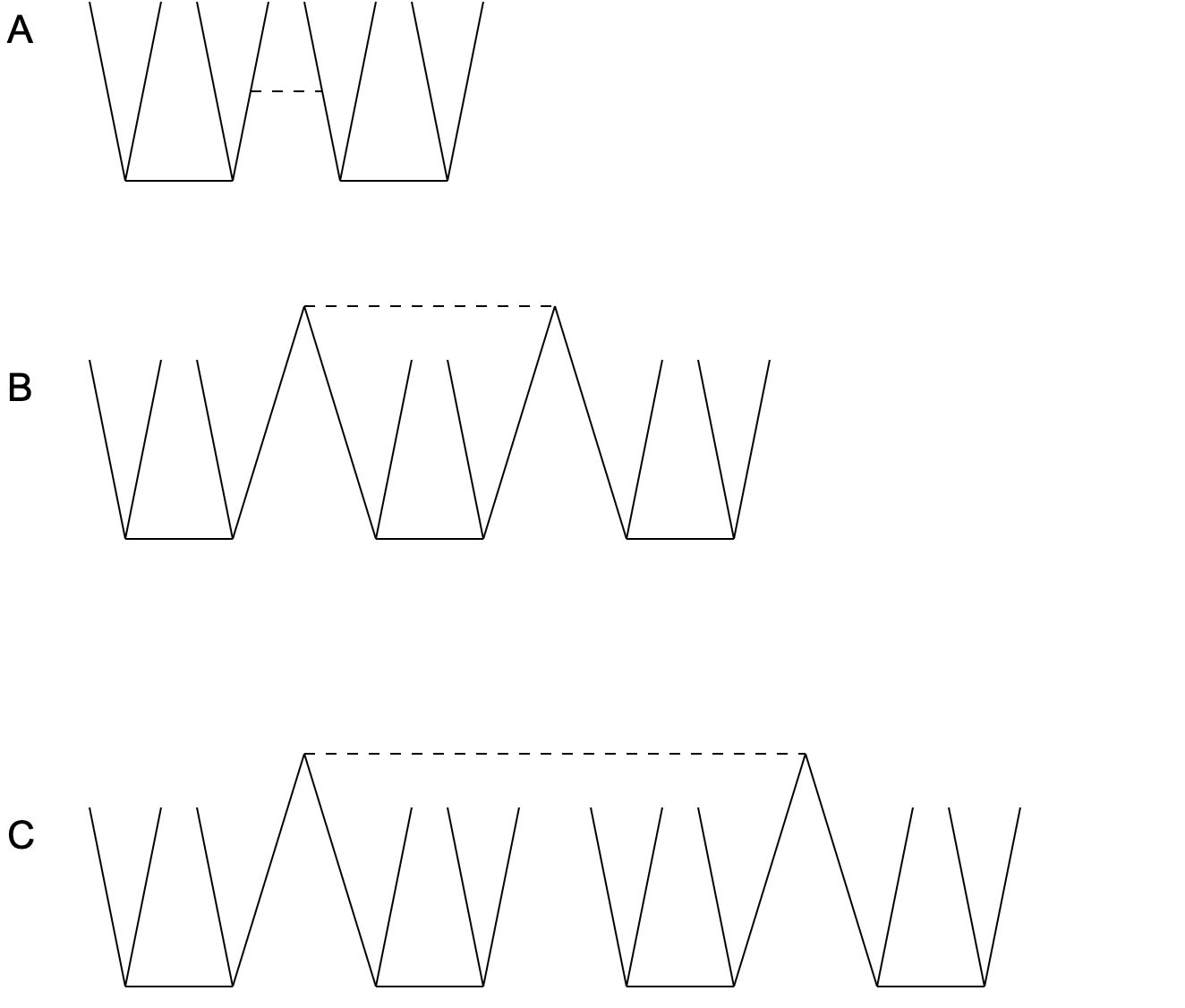}
\caption{Depiction of the bare-bone diagrams that will be used to construct higher order $T_{2n}$ cluster operators, at which point the pertinent resolvent lines are implied. In that case, A) pertains to $T_4^{[3]}$, B) pertains to $T_4^{[4]}$, and C)  pertains to $T_6^{[5]}$.}
\label{fig:setOfT2s}
\end{figure}

\subsubsection{$\Lambda$ motivations for PT-CCD(5) - PT-CCD(7)}
Let us specifically consider energy corrections through orders 5-7 of PT. We know these contributions will involve a $T_4$, correct through orders 3-5 in PT. To keep things general, the residual equations we are interested in solving are of the form
\begin{equation}\label{eq:fullT4residEqns}
    D_4T_4 =Q_4\bigg(\big(\frac{W_NT_2^2}{2}\big)_c + \big(\frac{W_NT_2^3}{3!} \big)_c + \big( W_NT_4T_2 \big)_c \bigg)\ket{0}
\end{equation}

However, we need to project this quantity into the space of doubly excited determinants so that an energy correction can be extracted. In other words, we need to contract the above $T_4$ with a $W_N^{-2}$ so that
\begin{equation}\label{eq:t2eqn}
\begin{split}
        D_2T_2=&W_N T_4\\
    \implies T_2(T_4)=&\frac{W_NT_4}{D_2}
\end{split}
\end{equation} where $T_2(T_4)$ represents a net-T2 amplitude that is constructed using a $T_4$. This can be subsequently  used to determine the energy correction 
\begin{equation} \label{eq:energyT4}
\begin{split}
        E(T_4) =& \braket{0|W_N T_2(4)|0}\\
               =& \braket{0|\frac{W_N W_N}{D_2D_4}D_4T_4|0}\\
\end{split}
\end{equation} where we have inserted Equation \ref{eq:t2eqn} and the identity $\frac{D_4}{D_4}$ into the above. Referring to Figure \ref{fig:factorizationTheorem} A, we note that there are 2 distinct time orderings for $D_2$ in the above equation such that $D_4^{AB}$ term can be factored out. By virtue of the factorization theorem, this leads to
\begin{equation}\label{eq:factorize2}
    \frac{1}{2}\frac{1}{D_4^{AB}}\bigg( \frac{1}{D_2^A} + \frac{1}{D_2^B}\bigg) \equiv \frac{1}{2D_2^A D_2^B}
\end{equation} Using Equation \ref{eq:factorize2} within Equation \ref{eq:energyT4}, we find that the energy can be rewritten in its factorized form as 
\begin{equation}\label{eq:finalT4E}
    E(T_4)=\frac{1}{2}\braket{0|\frac{W_NW_N}{D_2^AD_2^B}D_4T_4|0}
\end{equation} where we have explicitly removed the 8-index, $D_4$ tensor in the denominator. 

At this point, the remainder of the problem involves us 
determining the residual equations $D_4T_4$. Referring back to the RHS of Equation \ref{eq:fullT4residEqns}, the terms in sequential order correspond to third ($T_4^{[3]}$), fourth ($T_4^{[4]}$), and fifth-order ($T_4^{[5]}$) amplitudes. By inserting these individual residual equations into Equation \ref{eq:finalT4E}, the result is  factorized expressions for energies correct through fifth-order
\begin{equation}\label{eq:factorFifthOE}
    E(5)\equiv E(T_4^{[3]})=\frac{1}{2}\braket{0|\frac{W_NW_N}{D_2^AD_2^B}Q_4\bigg(\frac{W_NT_2^2}{2}\bigg)_c|0}
\end{equation} sixth-order:
\begin{equation}
   E(6)\equiv E (T_4^{[4]})=\frac{1}{2}\braket{0|\frac{W_NW_N}{D_2^AD_2^B}Q_4\bigg(\frac{W_NT_2^3}{3!}\bigg)_c|0} 
\end{equation} and seventh-order:
\begin{equation}\label{eq:seventhOET4}
    E(7)\equiv E(T_4^{[5]})=\frac{1}{2}\braket{0|\frac{W_NW_N}{D_2^AD_2^B}Q_4\bigg(\frac{W_NT_2T_4^{[3]}}{2!}\bigg)_c|0}
\end{equation} in PT. To be clear, Equation \ref{eq:seventhOET4} explicitly uses a $T_4^{[3]}$ given by the first term in Equation \ref{eq:fullT4residEqns}.

Alternatively, we arrive at the same results by expanding the CC functional of Equation \ref{eq:CCfunctional} and grouping equivalent-order terms. As an example, upon expanding to fifth-order energy contributions 
\begin{equation}\label{eq:fifthOrderELambda}
\begin{split}
E(5)=&\underbrace{\braket{0|\Lambda_4^{(2)}\bigg(H_0T_4^{[3]}\bigg)_c|0} + \braket{0|\Lambda_4^{(2)}\bigg(\frac{W_NT_2^2}{2}\bigg)_c|0}}_{\Lambda_4^{(2)} Q_4\big( H_0T_4^{[3]} + \frac{W_NT_2^2}{2} \big)\ket{0} \equiv 0} + \braket{0|\Lambda_2^{(1)}W_NT_4^{[3]}|0} \\
\implies &E(5)=\braket{0|\frac{W_NW_ND_4T_4^{[3]}}{D_2D_4}|0}
\end{split}
\end{equation} we see that the ROI can be inserted in the first two terms, leading to a set of residual equations that are 0 by construction. We note that in the last line, the form is similar to that found in Equation \ref{eq:energyT4}, just prior to invocation of the factorization theorem.  Equivalence between the final expression of Equation \ref{eq:fifthOrderELambda} and Equation \ref{eq:factorFifthOE} can be obtained by recognizing the expanded form of $\Lambda_2^{(1)}$, and invoking the factorization theorem as expressed by Equation \ref{eq:factorize2}. Upon doing so, we find that the final form gives us a factor of $\frac{W_N^AW_N^B}{D_2^AD_2^B}$, giving us the flexibility to rewrite the expression such that:
\begin{equation}
    E(5)=\frac{1}{2}\braket{0|\Lambda_4^{(2)}Q_4\bigg( \frac{W_NT_2^2}{2}\bigg)_c|0}
\end{equation} where $\Lambda_4^{(2)}=\Lambda_2^{(1)}\Lambda_2^{(1)}\equiv \frac{\braket{ij||ab}}{D_2^A}\frac{\braket{kl||cd}}{D_2^B} \equiv \frac{W_N}{D_2^A}\frac{W_N}{D_2^B}$. 

Expanding the CC functional to include sixth-order terms and inserting the ROI, we find
\begin{equation}
\begin{split}
E(6)=&\underbrace{\braket{0|\Lambda_4^{(2)}\big(H_0T_4^{[4]}\big)_c|0} +\braket{0|\Lambda_4^{(2)}\big(\frac{W_NT_2^3}{3!}\big)_c|0}}_{\Lambda_4^{(2)}Q_4\bigg( H_0T_4^{[4]} + \frac{W_NT_2^3}{3!}\bigg)_c\ket{0}=0} + \braket{0|\Lambda_2^{(1)}\bigg(W_NT_4^{[4]}\bigg)_c|0} \\
\implies & E(6) = \braket{0|\frac{W_NW_ND_4T_4^{[4]}}{D_2D_4}|0}
\end{split}
\end{equation} Following the same procedure as we did for $E(5)$, we find that this expression can be rewritten such that
\begin{equation}
    E(6)=\frac{1}{2}\braket{0|\Lambda_4^{(2)}\bigg(\frac{W_NT_2^3}{3!}\bigg)_c|0}
\end{equation}

Similarly, expanding the CC functional to seventh-order results in
\begin{equation}
    E(7) = \braket{0|\Lambda_2^{(1)}\bigg( W_NT_4^{[5]}\bigg)_c|0}
\end{equation}after eliminating the terms present in the  $Q_4 T_6^{[5]}$ residual equations. Again, this can be rewritten as
\begin{equation}
    E(7)=\frac{1}{2}\braket{0|\Lambda_4^{(2)}\bigg(W_NT_4^{(3)}T_2\bigg)_c|0}
\end{equation}

It is important to note that at this point, the fifth and sixth order expressions are exactly those prescribed by the XCCD energy functional at corresponding orders. However, the seventh order energy correction extracted by the $\Lambda$ formulation is notably distinct from that in XCCD, as the former explicitly makes use of $T_4^{(3)} \equiv \frac{(W_nT_2^2)_C}{D_4}$ whereas the latter only makes use of $(W_nT_2^2)_C$, forgoing any use of the higher-order $D_4$ denominator. The differences betweent the XCCD energy expressions and those generated by the above $\Lambda$ formulation are a byproduct in how the energy is generated. For $\Lambda$, we strictly rely on the residual equantion to be connected before the energy functional is evalutated. In XCCD, the energy functional is constructed first with the only requirement being that the contributing diagrams -arising from products on $T_2$ only-are connected \textit{en toto}. Consequently, the residual equations arising from the XCCD energy functional do not explicitly ``see" any denominator $D_{2n}$ for $n>1$.

\subsubsection{$\Lambda$ motivations for PT-CCD(8/9)}
Beginning with eighth-order energy corrections, $T_6$ becomes relevant. For convenience, we follow the CC functional route to invoke the factorization theorem from this point forward. Noting that if we use the $Q_6T_6^{[5]}$ and $Q_4T_4^{[6]}$ residual equations to eliminate terms, we are left with 
\begin{equation}\label{eq:eigthOE}
\begin{split}
        E(8)=&\frac{1}{2}\braket{0|\Lambda_4^{(2)}W_NT_6^{[5]}|0} \\
        =& \frac{1}{2}\braket{0|\frac{W_NW_N}{D_2^AD_2^B}\frac{W_N}{D_6^{ABC}}D_6T_6^{[5]}|0}
\end{split}
\end{equation} Using Figure \ref{fig:factorizationTheorem} B as guidance, we see that
\begin{equation}\label{eq:D6factorize}
    \frac{1}{3}\frac{1}{D_6^{ABC}}\bigg( \frac{1}{D_2^AD_2^B} + \frac{1}{D_2^AD_2^C} + \frac{1}{D_2^BD_2^C}\bigg) = \frac{1}{3}\frac{1}{D_2^AD_2^BD_2^C}
\end{equation} using the fact that after we set equivalent denominators, $D_2^A + D_2^B + D_2^C =D_6^{ABC}$. Inserting Equation \ref{eq:D6factorize} into Equation \ref{eq:eigthOE} yields
\begin{equation}
\begin{split}
    E(8)&=\frac{1}{3!}\braket{0|\Lambda_6^{(3)}D_6T_6^{[5]}|0}\\
    &=\frac{1}{3!}\braket{0|\Lambda_6^{(3)}\big(\frac{W_NT_2^4}{4!}\big)_C|0}
\end{split}
\end{equation}.

Similarly for the ninth-order energy expression, we can use the $D_6T_6^{[6]}$ residual equation to eliminate terms which leaves
\begin{equation}
    E(9)=\frac{1}{2}\braket{0|\Lambda_4^{(2)}W_NT_6^{[6]}|0}
\end{equation}Following the above prescription, we find that the rewritten ninth-order energy expression can be rewritten such that
\begin{equation}
\begin{split}
    E(9)&=\frac{1}{3!}\braket{0|\Lambda_6^{(3)}D_6T_6^{[6]}|0}\\
    &=\frac{1}{3!}\braket{0|\Lambda_6^{(3)}\big(W_NT_4^{[3]}T_2\big)_C|0}
    \end{split}
\end{equation}
Note that the $\Lambda$ expression for $E(8)$ is exactly equivalent to the energy expression found in the XCCD(8) energy functional, but is the last time these methods will exactly coincide. Evidence of this becomes clear for $E(9)$, where the $\Lambda$ formulation explicitly builds a $T_4^{[3]}$ which involves $D_4$ whereas XCCD(9) simply uses $\frac{W_NT_2^2}{2}$ in its stead and omits use of $D_4$.

\section*{References}

\bibliography{main.bib}

\end{document}